\documentclass{aastex631}
\usepackage{subfigure}
\usepackage{amsmath}
\usepackage{amsmath,amsthm,amssymb,amsfonts}
\usepackage{multirow}
\usepackage{verbatim}
\usepackage{blindtext}
\usepackage{bm}

\usepackage{textcomp}
\usepackage{hyperref}
\usepackage[misc]{ifsym} 

\graphicspath{{./}{figures/}}

\begin{document}

\title{Multibeam Blind Search of Targeted SETI Observations toward 33 Exoplanet Systems with FAST}

\author[0000-0003-3977-4276]{Xiao-Hang Luan}
\affiliation{Institute for Frontiers in Astronomy and Astrophysics, Beijing Normal University, Beijing 102206, People's Republic of China}
\affiliation{Department of Astronomy, Beijing Normal University, Beijing 100875, People's Republic of China; \url{tjzhang@bnu.edu.cn}}

\author[0000-0002-4683-5500]{Zhen-Zhao Tao}
\affiliation{Institute for Frontiers in Astronomy and Astrophysics, Beijing Normal University, Beijing 102206, People's Republic of China}
\affiliation{Department of Astronomy, Beijing Normal University, Beijing 100875, People's Republic of China; \url{tjzhang@bnu.edu.cn}}

\author[0000-0002-5485-1877]{Hai-Chen Zhao}
\affiliation{Institute for Frontiers in Astronomy and Astrophysics, Beijing Normal University, Beijing 102206, People's Republic of China}
\affiliation{Department of Astronomy, Beijing Normal University, Beijing 100875, People's Republic of China; \url{tjzhang@bnu.edu.cn}}

\author{Bo-Lun Huang}
\affiliation{Institute for Frontiers in Astronomy and Astrophysics, Beijing Normal University, Beijing 102206, People's Republic of China}
\affiliation{Department of Astronomy, Beijing Normal University, Beijing 100875, People's Republic of China; \url{tjzhang@bnu.edu.cn}}
\affiliation{School of Physics and Astronomy, University of Glasgow, Glasgow G12 8QQ, United Kingdom}

\author{Shi-Yu Li}
\affiliation{Beijing Planetarium, Beijing Academy of Science and Technology, Beijing 100044, People's Republic of China}

\author{Cong Liu}
\affiliation{Institute for Astronomical Science, Dezhou University, Dezhou 253023, People's Republic of China}

\author{Hong-Feng Wang}
\affiliation{Institute for Astronomical Science, Dezhou University, Dezhou 253023, People's Republic of China}

\author{Wen-Fei Liu} 
\affiliation{College of Physics and Electronic Engineering, Qilu Normal University, Jinan 250200, People's Republic of China}

\author[0000-0002-3363-9965]{Tong-Jie Zhang\href{mailto:tjzhang@bnu.edu.cn}{\textrm{\Letter}}}
\affiliation{Institute for Frontiers in Astronomy and Astrophysics, Beijing Normal University, Beijing 102206, People's Republic of China}
\affiliation{Department of Astronomy, Beijing Normal University, Beijing 100875, People's Republic of China; \url{tjzhang@bnu.edu.cn}}
\affiliation{Institute for Astronomical Science, Dezhou University, Dezhou 253023, People's Republic of China}

\author[0000-0002-8604-106X]{Vishal Gajjar}
\affiliation{Breakthrough Listen, University of California Berkeley, Berkeley, CA 94720, USA; \url{vishalg@berkeley.edu}}

\author{Dan Werthimer}
\affiliation{Breakthrough Listen, University of California Berkeley, Berkeley, CA 94720, USA}
\affiliation{Space Sciences Laboratory, University of California Berkeley, Berkeley, CA 94720, USA}


\begin{abstract}

The search for extraterrestrial intelligence (SETI) is to search for technosignatures associated with extraterrestrial life, such as engineered radio signals. In this paper, we apply the multibeam coincidence matching (MBCM) strategy, and propose a new search mode based on the MBCM which we call the MBCM blind search mode. In our recent targeted SETI research, 33 exoplanet systems are observed by the Five-hundred-meter Aperture Spherical radio Telescope (FAST). With this blind search mode, we search for narrowband drifting signals across $1.05-1.45$ GHz in two orthogonal linear polarization directions separately. There are two special signals, one of which can only be detected by the blind search mode while the other can be found by both blind and targeted search modes. This result reveals huge advantages of the new blind search mode. However, we eliminate the possibility of the special signals being ETI signals based on much evidence, such as the polarization, drift, frequency and beam coverage characteristics. Our observations achieve an unprecedented sensitivity and our work provides a deeper understanding to the polarization analysis of extraterrestrial signals.



\end{abstract}

\keywords{\href{http://astrothesaurus.org/uat/74}{Astrobiology (74)}; \href{http://astrothesaurus.org/uat/2127}{Search for extraterrestrial intelligence (2127)}; \href{http://astrothesaurus.org/uat/2128}{Technosignatures (2128)}; \href{http://astrothesaurus.org/uat/498}{Exoplanets (498)}}


\section{Introduction} \label{sec:intro}

The question of whether or not the Earth is the only host for life in the universe has long fascinated human beings. Based on the Copernican principle and Drake equation \citep{1961PhT....14d..40D}, most scientists believe that there must be intelligent life beyond Earth (i.e. Extraterrestrial Intelligence, or ETI for short). In addition to theoretical research, a series of recent discoveries show that Earth-like exoplanets exist in large numbers throughout the Milky Way \citep{2013ApJ...767...95D,2013PNAS..11019273P,2014PNAS..11112647B}.

To search for life beyond Earth, there are three main approaches: (1) direct in situ detection of biosignatures (such as by-products of life or biological processes) in specific environments \citep{2015Sci...347..415W}; (2) remote sensing biological signals from the atmospheres and surfaces of exoplanets \citep{2014Sci...343..171R,2014PNAS..11112634S}; and (3) detection of technosignatures from extraterrestrial intelligence i.e. SETI \citep{1959Natur.184..844C}. From an astronomical perspective, it is extremely challenging to detect the biosignatures, due to the limitations of detection range and observation time \citep{2005AsBio...5..706S,2014ApJ...781...54R,2016ApJ...819L..13S}. In contrast, SETI only needs to probe the technological signals intentionally or unintentionally produced by ETI. Most SETI experiments are usually conducted in the radio wave band, given the effectiveness of radio signal propagation through interstellar space. Early Radio SETI observations only searched at specific frequencies, such as the hydrogen hyperfine transition 21 cm line \citep{1959Natur.184..844C} and the hydroxyl 18 cm lines \citep{1980Icar...42..136T}. With the advancement of radio instrumentation, the available bandwidth of SETI systems has expanded to tens of GHz \citep{2018PASP..130d4502M}.

The narrowest radio spectrum of natural astrophysical phenomena has a width of at least $\sim$ 500 Hz, because EM signals in nature experience broadening effects \citep{1987MNRAS.225..491C}. Thus, narrow bands ($\sim$ Hz) are well suited in terms of carrying information for ETI \citep{2001ARA&A..39..511T}. A narrowband signal transmitted from a distant source shows Doppler drift in frequency due to the motion of the transmitter relative to the receiver. Therefore, narrowband drifting signals are targets for SETI research.

Radio SETI observations can be carried out in two ways, sky survey and targeted search \citep{2001ARA&A..39..511T}. Sky surveys are mainly represented by projects such as SERENDIP, SETI@home \citep{2001SPIE.4273..104W}, and FAST SETI backend \citep{2020ApJ...891..174Z}. In recent years, targeted SETI observations are increasingly being conducted toward exoplanets and some nearby stars by many radio telescopes in various countries. For example: 
studies from the Breakthrough Listen initiative (BL) using the Green Bank Telescope (GBT) and the Parkes Telescope (Parkes) \citep{2017ApJ...849..104E,2020AJ....159...86P,2020AJ....160...29S,2021AJ....161..286T,2021NatAs...5.1148S,Gajjar_2021}; observations toward planetary systems by the Allen Telescope Array (ATA) \citep{2016AJ....152..181H,2018ApJ...869...66H,2020AJ....160..162H}; many SETI surveys using the Murchison Widefield Array (MWA) \citep{2016ApJ...827L..22T,2018ApJ...856...31T,2020PASA...37...35T}; observations focused on Sun-like Stars and planetary systems in the Kepler field with GBT \citep{2019AJ....157..122P,2021AJ....161...55M}; a Very Large Array (VLA) search for technosignatures from M31 and M33 \citep{2017AJ....153..110G}; and the targeted observations of 33 exoplanets by China's Five-hundred-meter Aperture Spherical radio Telescope (FAST) \citep{Tao_2022}. However, there is no conclusive evidence of ETI technosignatures from any research so far.

The major challenge of Radio SETI research is identifying radio frequency interference (RFI) caused by human technology. To reject RFI, one of the techniques employed by BL is called the on-off strategy \citep{2013ApJ...767...94S,2017ApJ...849..104E,2019AJ....157..122P,2020AJ....159...86P,2020AJ....160...29S,2021AJ....161..286T,2021NatAs...5.1148S,Gajjar_2021}, which is the conventional method to identify RFI in targeted SETI observations. On-observation is observing right at a selected target. Off-observation is to observe positions near the target as reference observations, which are at least several times the full width at half maximum of the telescope from the target. An ETI signal from the target should only appear in the on-observation but not in any off-observations. However, most RFI can be detected in both on and off observations due to its large radiation angle. The on–off strategy has been applied to GBT and Parkes so far.

FAST \citep{2000ASPC..213..523N,2011IJMPD..20..989N,2016RaSc...51.1060L,2019SCPMA..6259502J,2020RAA....20...64J}, the largest single-aperture radio telescope on Earth, has an L-band ($1.05-1.45$ GHz) 19-beam receiver, a large sky area coverage (declinations from $-14^{\circ}.34$ to $+65^{\circ}.7$), and extremely high sensitivity ($\sim 2000\;m^{2}\; K^{-1}$, Arecibo’s is about $1100\; m^{2} \;K^{-1}$). Based on the above characteristics, FAST has a unique advantage in SETI observations \citep{2006ScChG..49..129N,2020RAA....20...78L}. In 2019, the first SETI sky survey of FAST was conducted by commensal observations \citep{2020ApJ...891..174Z}. In 2021, the first targeted SETI observations were also conducted by FAST toward exoplanet systems \citep{Tao_2022}. Since the 19-beam receiver can simultaneously receive signals from different sky regions, the targeted SETI observations of FAST can identify RFI without using the on-off strategy. \citet{Tao_2022} designed an effective observation strategy named multibeam coincidence matching (MBCM) for FAST SETI observations.

In this paper, we propose a new search mode in targeted SETI observations, called MBCM blind search mode. Moreover, we report the results of observations of 33 exoplanet systems after applying MBCM blind search mode. In Section \ref{sec:observe}, we discuss the targets, the criteria of the MBCM blind search mode, and the FAST multibeam digital backend. The data analysis techniques are discussed in Section \ref{sec:analysis}. The results of signal search and identification are presented in Section \ref{sec:results}. In Section \ref{sec:discussion and conclusions}, we briefly summarize this work and discuss the advantages of the MBCM blind search mode, the understanding of extraterrestrial signal polarization analysis, and the particularity of two signals of interest.

\section{Observations} \label{sec:observe}

\begin{figure*}[!htp]
	
	\centering
	\subfigure[]{
		\includegraphics[width=0.29\linewidth]{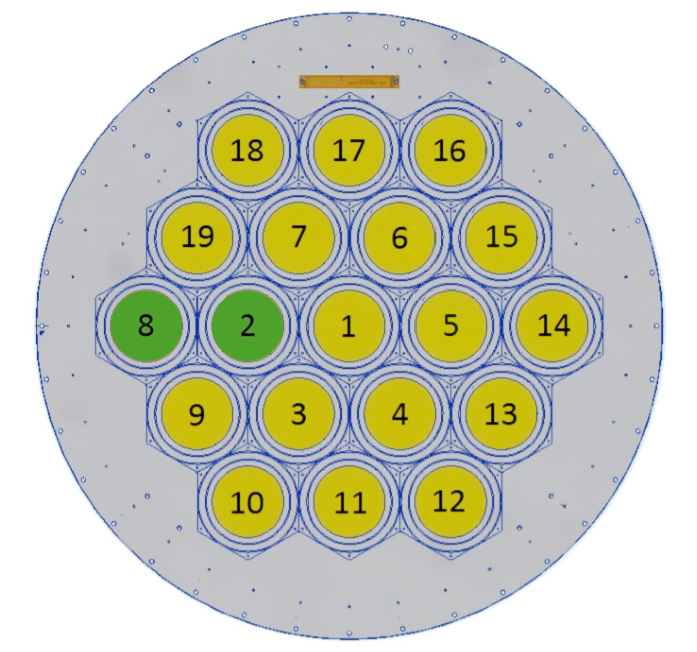}
	}
	\quad
	\subfigure[]{
		\includegraphics[width=0.29\linewidth]{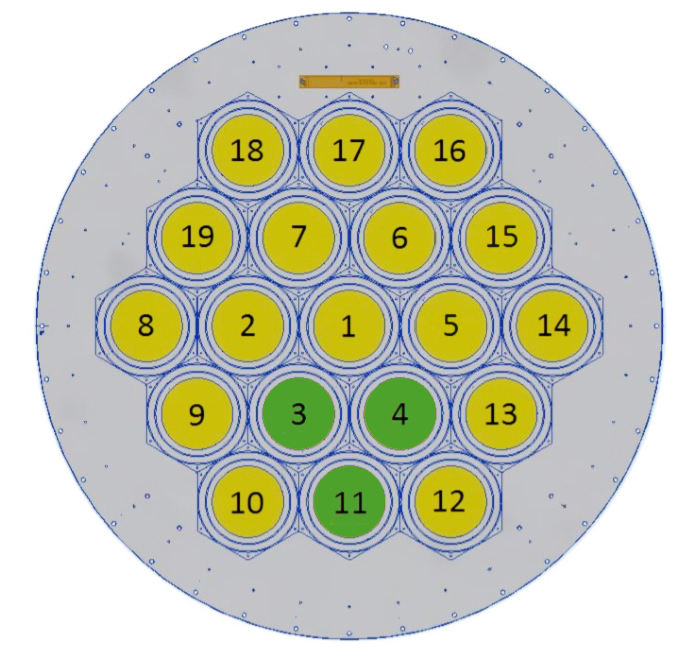}
	}
	\quad
	\subfigure[]{
		\includegraphics[width=0.29\linewidth]{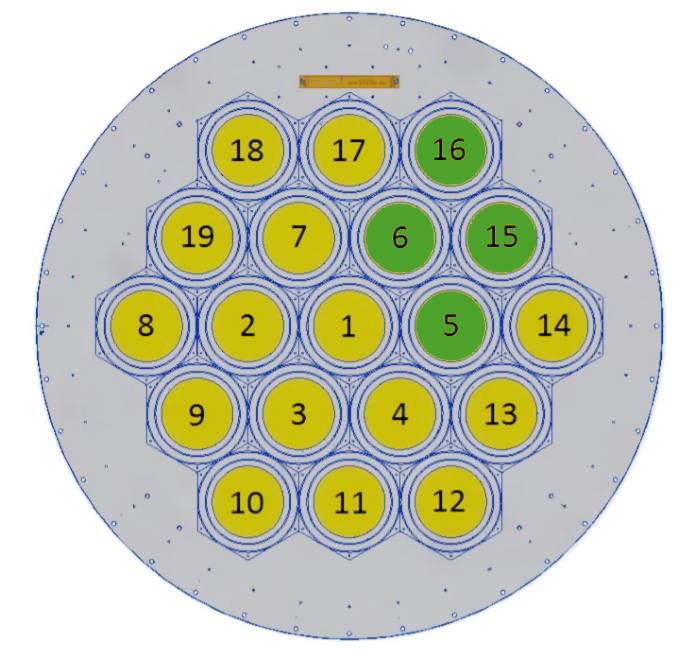}
	}
	
	\subfigure[]{
		\includegraphics[width=0.29\linewidth]{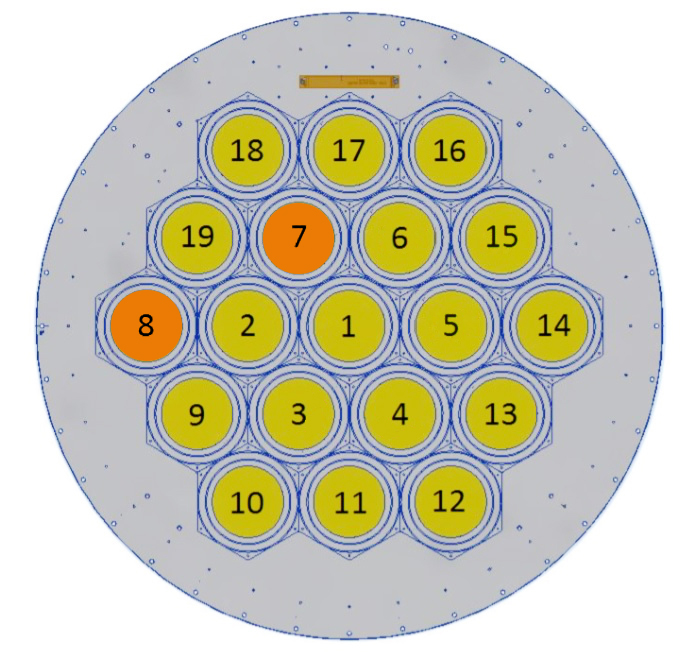}
	}
	\quad
	\subfigure[]{
		\includegraphics[width=0.29\linewidth]{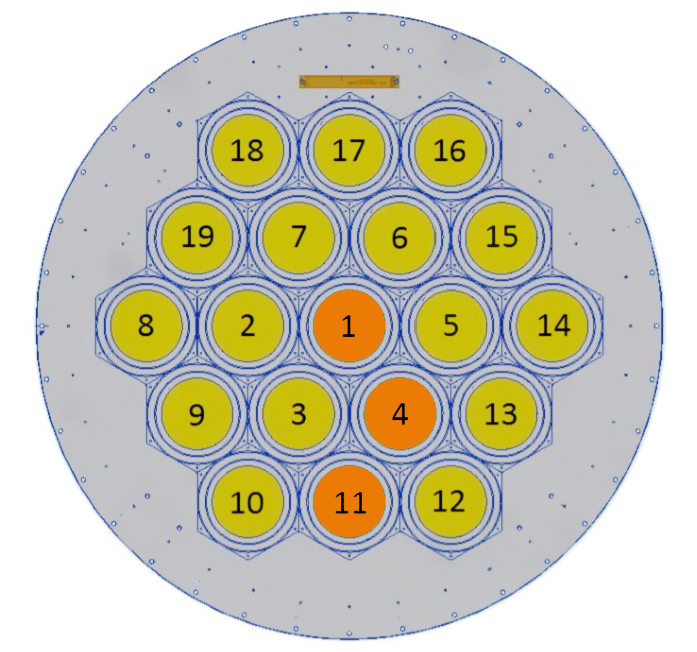}
	}
	\quad
	\subfigure[]{
		\includegraphics[width=0.29\linewidth]{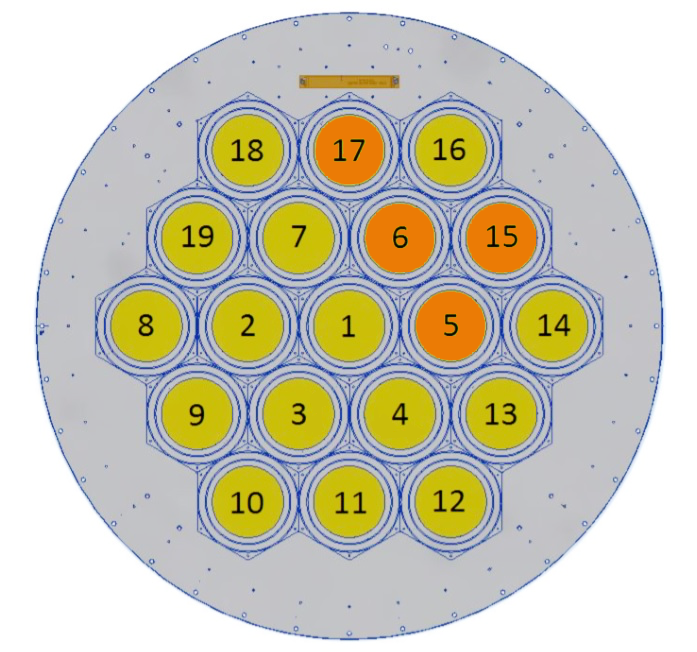}
	}
	\caption{Schematics of the MBCM blind search mode. The first row shows three examples of allowed signals, and the bottom row shows three examples of forbidden signals.}
	\label{fig:beams}
\end{figure*}


\subsection{Targets} \label{subsec:Targets}

In SETI research, habitable zone \citep{KASTING1993108,williams_pollard_2002,Kopparapu_2013} and earth transit zone \citep{2021Natur.594..505K} are considered some of the perfect places to look for ETI signals. Habitable zone is traditionally defined as the region around a star in which a terrestrial-mass planet with a $\mathrm{CO}_{2}\mathrm{H}_{2}\mathrm{ON}_{2}$ atmosphere can maintain liquid water on its surface \citep{KASTING1993108}. The Earth transit zone is a region where observers outside the solar system can see the Earth transit the Sun. The source selection mainly takes into account the above two aspects, that is, we observe worlds that is similar to ours and worlds that can see us.

Referring to the data up to October 2020 from the NASA Exoplanet Archive \footnote{\url{https://exoplanetarchive.ipac.caltech.edu/}} and considering the observable sky of FAST, we observe 33 exoplanet systems from April 19, 2021 to September 21, 2021. More importantly, the targets are selected considering habitability, celestial position, size, and distance, 29 out of 33 exoplanet systems were detected to contain planets in their habitable zones, and 5 of them are within the earth transit zone \citep{Tao_2022}.


\subsection{Strategy} \label{subsec:strategy}

The basic principle of the on-off strategy is that an ETI signal from the target can only be detected in the on-observation, while RFI can be detected in both on and off observations \citep{2017ApJ...849..104E}. Based on the FAST 19-beam receiver, \citet{Tao_2022} designed the MBCM strategy for targeted SETI observations. The MBCM strategy states that multiple beams are observed and compared simultaneously to identify the origin of a signal. Through different beam combinations, the MBCM strategy can be conducted in two modes so far: targeted search mode \citep{Tao_2022} and blind search mode. MBCM targeted search mode corresponds to the on-off strategy. Beam 1 (central beam) serves as the on-observation, and the six outermost beams (Beam 8, 10, 12, 14, 16, and 18) serve as the off-observations (i.e. a total of seven beams). To use all 19 beams to search for ETI signals, we propose the new MBCM blind search mode in this paper, which can detect signals within the 19-beam-pointed sky ($\sim$ $11.6^{\prime}$) around the targets. Beam 1 (central beam) keeps tracking the target, and all other beams serve as reference beams.

The idea of MBCM blind search mode comes from the multibeam blind search for FRBs \citep{10.1093/mnras/stx2126,2019SSPMA..49i9508P,Zhu_2020}. The layout of the 19 beams on the FAST L-band receiver is shown in Figure \ref{fig:beams}. We classify a signal as RFI if the signal is covering non-adjacent beams, or covering more than four adjacent beams, or covering three or more beams in a line. When non-adjacent beams detect the same signal with obvious directivity, the intensity attenuation is at least 30 dB or more due to the design of the receiver, which makes the signal hardly detectable. However, the radiation angle of RFI is generally large so that multiple beams can detect signals above the set threshold. When a signal is detected by the multibeam receiver, it can be detected by one or more beams, and the combination of beams which detect the signal is called a beam coverage arrangement. Furthermore, beam coverage arrangement from a signal that could be potentially transmitted by an ETI is called an allowed beam coverage arrangement. There are four types of allowed beam coverage arrangement for the MBCM blind search mode: (1) one beam, could be any of the 19 beams; (2) two adjacent beams, such as Beams 2 and 8 (Figure \ref{fig:beams}(a)); (3) three beams adjacent to each other forming an equilateral triangle, such as Beams 3, 4, and 11 (Figure \ref{fig:beams}(b)); (4) four beams adjacent to each other forming a compact rhombus, such as Beams 5, 6, 15, and 16 (Figure \ref{fig:beams}(c)).

However, any beam coverage arrangements other than the four types listed above are called forbidden beam coverage arrangements, such as the three examples in the second row shown in Figure \ref{fig:beams}: (1) a signal detected by only Beams 7 and 8 (Figure \ref{fig:beams}(d)). The angular distance between Beam 7 and Beam 8 is about the width of one beam, if an extraterrestrial signal is between Beam 7 and Beam 8, Beam 2 and Beam 19 should have a stronger response as they are closer; (2) a signal detected only in Beams 1, 4, and 11 (Figure \ref{fig:beams}(e)) is also forbidden. Similar to (1), if these three beams can all detect the signal, Beam 3 can also detect it because of shorter angular distance; and (3) a signal detected only in Beams 5, 6, 15, and 17 (Figure \ref{fig:beams}(f)) in this beam coverage arrangement, since Beams 5, 6, and 17 are arranged in a line, an ETI signal is unlikely to cover them simultaneously. All signals with the above three similar beam coverage arrangements are likely to be RFI. Other cases where beams are farther away or more beams are arranged in a line are also considered as RFI, such as Beams 4 and 9 or Beams 9, 3, 4, and 13. The maximum number of beams in an allowed beam coverage arrangement is set to four, because when the number of beams reaches five, at least three beams are arranged in a line.


The effectiveness of the MBCM blind search mode can be quantitatively analyzed by the beam response in the 19-beam receiver. The angular distance between two adjacent beams in the 19-beam receiver is $5.8^{\prime}$. For a uniformly illuminated circular aperture, the distribution of the beam response $I$ with the angular distance from the beam centre $\theta$ is given by
\begin{equation}
	I(\theta)=I_{0}\left[\frac{2 J_{1}(k a \sin \theta)}{k a \sin \theta}\right]^{2},
\end{equation} 
where $I_{0}$ is the intensity at the beam centre, $k=2 \pi / \lambda$ is the wave number, $a$ is the radius of the illuminated aperture, and $J_{1}$ is the Bessel function of the first kind of order one. This expression is also known as the Airy pattern. The maximum of the first side lobe occurs at
\begin{equation}
	\theta_{1}=1.63 \frac{\lambda}{D},
\end{equation}
where $\theta_{1}$ is in radians and $D=2 a$. The response there is $I\left(\theta_{1}\right)=1.75 \% I_{0}$, which is already a very small value. The responses of the other side lobes are weaker and can generally be ignored. In uniform illumination, the maximum of the first side lobe decreases with the increase of frequency. Given the effective frequency range ($1.05-1.45$ GHz) of the L-band receiver and a = 150 m for FAST, the maximum of the first side lobe is located in $\theta_{1}=5.34^{\prime}$ at 1.05 GHz, which is the largest angular distance. The measurement of the FWHM beamwidth of the FAST L-band 19-beam receiver indicates that the illumination pattern of FAST is between the uniform and cosine-tapered illumination \citep{2019SCPMA..6259502J}. A cosine-tapered illumination increases the FWHM beam width of the main lobe and decreases the side lobe level so that the maximum of its first side lobe is smaller.

ETI signals are generally thought to have strong directionality. The central beam (Beam 1) is most likely to detect the signal because it points exactly at the target, but it does not rule out the possibility that signals can also appear in other beams or the mid-point of adjacent beams.


In this example, our discussion is based on the condition that the signal is neither extremely faint nor extremely bright. When a signal is originated at the sky location where is not pointed by Beam 1 or Beam 2 but exactly their mid-point, the angular distance between the signal and the centres of these two beams is $2.9^{\prime}$, which is less than $5.34^{\prime}$. Thus, Beams 1 and 2 are able to detect the signal. In addition, Beams 3 and 7 are $5.02^{\prime}$ away from the signal, so there is a higher chance for them to detect it. Thus, these four beams arranged in a compact rhombus shape are all more likely to detect a signal that originated at geometric centre of the rhombus, while other beams are less likely to detect it because the angular distances between the signal and the beams outside the rhombus are well above $5.34^{\prime}$. Similarly, as shown in Figure \ref{fig:beams}(b), suppose an ETI signal is right at the centre of the equilateral triangle that consists of three beams adjacent to each other. In this case, the signal is visible for all three beams since the angular distances ($\sim3.35^{\prime}$) between the three beams and the signal are all under $5.34^{\prime}$.

In addition, we decided not to consider signals that were detected by more than four beams as ETI candidates. In practise, signals detected by only one beam have all outnumbered signals detected by two, three or four beams. Although counting signals covering more than four beams add extra beam coverage arrangements, the computational cost is going to be significantly higher. Therefore, the tradeoff point is set such that signals detected by more than four beams are not taken into account.

During our observations, we use all 19 beams in the L-band receiver to record data simultaneously. Beam 1 keeps tracking the target. Each observation lasts 20 minutes (except HD-111998, which lasts $\sim$ 4 minutes). Any detection with S/N above the threshold is then categorized as $hits$. Among all $hits$, we refer to $hits$ which meet the criteria of the MBCM blind search mode as $events$; otherwise they are classified as RFI. Using all 19 beams, we can detect and identify signals more effectively and comprehensively.

\subsection{FAST Multibeam Digital Backend} \label{subsec:hardware}

The hardware for digital signal sampling in the FAST multibeam digital backend consists of the Roach2 Field Programmable Gate Array (FPGA) board and the analogue-to-digital converter KatADC (hereafter ADC), which are plugged together and placed in the Roach2 chassis. Roach2 performs polyphase filtering, Fast Fourier Transform (FFT), correlation, integration, network encapsulation, and other processing for the sampled signal. Then it outputs the original voltage and power spectrum data through eight 10 Gb network ports. The ADC sampling frequency is 1 GHz, and the FPGA operating frequency is 250 MHz.

The signals of 19 beams are digitized by 10 Roach2 boards. In Roach2 1 – 9, each Roach2 processes two beam data, and Roach2 10 only processes one beam data (Figure 4 of \citet{Tao_2022}). Before being processed by Roach2, the signals in $X$ and $Y$ polarization are transmitted and sampled, respectively. After auto-correlation and cross-correlation of the two polarization data on Roach2, the output data includes four polarization channels: $XX$, $YY$, $XY$, $YX$. Since ETI signals may be narrowband drifting signals and persist for a long time, we record data on the spectral line backend, setting the frequency resolution to $\sim7.5$ Hz and the integration time to 10 seconds per spectrum.

\section{Data Analysis} \label{sec:analysis}

We record our data across $1.0-1.5$ GHz on the the spectral line backend using the L-band 19-beam receiver. Each FITS file records four polarization channels ($XX$, $YY$, $XY$, $YX$) of two spectra for each beam. The observation for each target lasts about 20 minutes  (except HD-111998), and the total data volume is 66.5 TB (including experimental observations). Before searching for signals, we first merge the FITS files of the same beam observed from the same target and then convert each merged FITS file into two Filterbank files recording the time-frequency two-dimensional power spectrum of $XX$ and $YY$ respectively. Filterbank files can be opened by the Blimpy package \citep{Price2019} for subsequent processing.

Targeted SETI in recent years mostly search only Stokes I data for narrowband signals \citep{2017AJ....153..110G,2020AJ....159...86P,2020AJ....160...29S,2021NatAs...5.1148S}. We process the intensity data of $XX$ and $YY$ separately to display the polarization characteristics and identify RFI generated by the backend instrument. We use the TurboSETI package \citep{2017ApJ...849..104E,2019ascl.soft06006E} to search for $hits$ in each beam. TurboSETI is a Python-Cython package using the tree search algorithm \citep{Siemion_2013} to search for narrowband signals from the time-frequency two-dimensional power spectrum. It breaks the limitation of time-frequency resolution on the maximum drift rate through a shifting array so that it can search for narrowband signals with arbitrary drift rates. TurboSETI opens the Filterbank file by calling the Blimpy package. Two key parameters in the TurboSETI algorithm are S/N threshold and the maximum drift rate (MDR), which enable it to search for narrowband signals above a S/N threshold, and drift rates between $\pm$ MDR. Setting a S/N threshold that is too high could result in filtering out most important signals and a threshold that is too low will usually fail to reject noise. Referring to previous targeted SETI studies using TurboSETI \citep{2017ApJ...849..104E,2020AJ....159...86P,2020AJ....160...29S,2021AJ....161..286T,2021NatAs...5.1148S,Gajjar_2021,Tao_2022}, we set the S/N threshold to 10. The drift rate of a signal is mainly caused by the rotation and orbit of the Earth and the target planet. According to \citet{Tao_2022}, the value of MDR depends on the observed frequency band because the drift rate is proportional to transmitted frequency of the signal. For FAST L-band, we set the MDR to 4 Hz/s, which is large enough to search for signals transmitted from a typical planet in habitable zone \citep{2022ApJ...938....1L,Tao_2022}. TurboSETI outputs the best fit frequency, drift rate, and $\mathrm{S} / \mathrm{N}$ of the $hits$ to a DAT file. Apart from using TurboSETI, we also use a de-drifting algorithm to calculate the drift rate with higher accuracy for the specific signal \citep{Tao_2022}.

After acquiring $hits$ using TurboSETI, the 19 beams are used to search for $events$. First, we list valid masks of all possible ETI signal cases according to the MBCM blind search mode criteria. This mask consists of 19 bits binary numbers, with `0' and `1' representing `mask' and `allow', respectively. The valid mask list contains: (1) all cases detected by a single beam, the valid masks are from 0000000000000000001 to 1000000000000000000, `1' is from the lowest bit to the highest bit; (2) all cases detected by two adjacent beams, such as Beams 2 and 8, the corresponding valid mask is 0000000000010000010; (3) all cases detected by three adjacent beams forming an equilateral triangle, such as Beams 3, 4, and 11, the corresponding valid mask is 0000000010000001100; and (4) all cases detected by four adjacent beams forming a compact rhombus, such as Beams 5, 6, 15, and 16, the corresponding valid mask is 0001100000000110000.

During data processing, we use 19 bit binary numbers with `0' and `1' to identify the received signal, called the identification code. We set the frequency error range to $\pm 5 \delta v$ ($\delta v$ is the frequency resolution). That is, the $hits$ within a frequency error range can be considered the same frequency and marked as `1', otherwise marked as `0'. First, we count the number of `1's in the identification code. If the number is more than four, add an RFI mark. If the number of `1's is less than or equal to four, we perform the `same-or operation' with the identification code and valid mask list. The `same-or' is a mathematical operator. If two values are the same, the result is 1, otherwise is 0. In the above operation, if the operation result is 1, add an ETI candidate mark and end this operation. If the result of the identification code with all valid masks is 0, add an RFI mark. For example, if the identification code is 1000000000110000010, which can match a valid mask, this signal is marked as an ETI candidate. For another example, if the identification code is 0000000101000001101, it is rejected as RFI because the number of `1's exceeds four.

The multibeam data processing of the MBCM blind search mode outputs $events$ of each target to a CSV file. The effective band of the L-band receiver is $1.05-1.45$ GHz \citep{2011IJMPD..20..989N}, as we remove the $events$ in a 50 MHz wide invalid band at both ends after processing.

\begin{figure*}[!htp]
	
	\centering
	\subfigure[]{
		\includegraphics[width=0.98\linewidth]{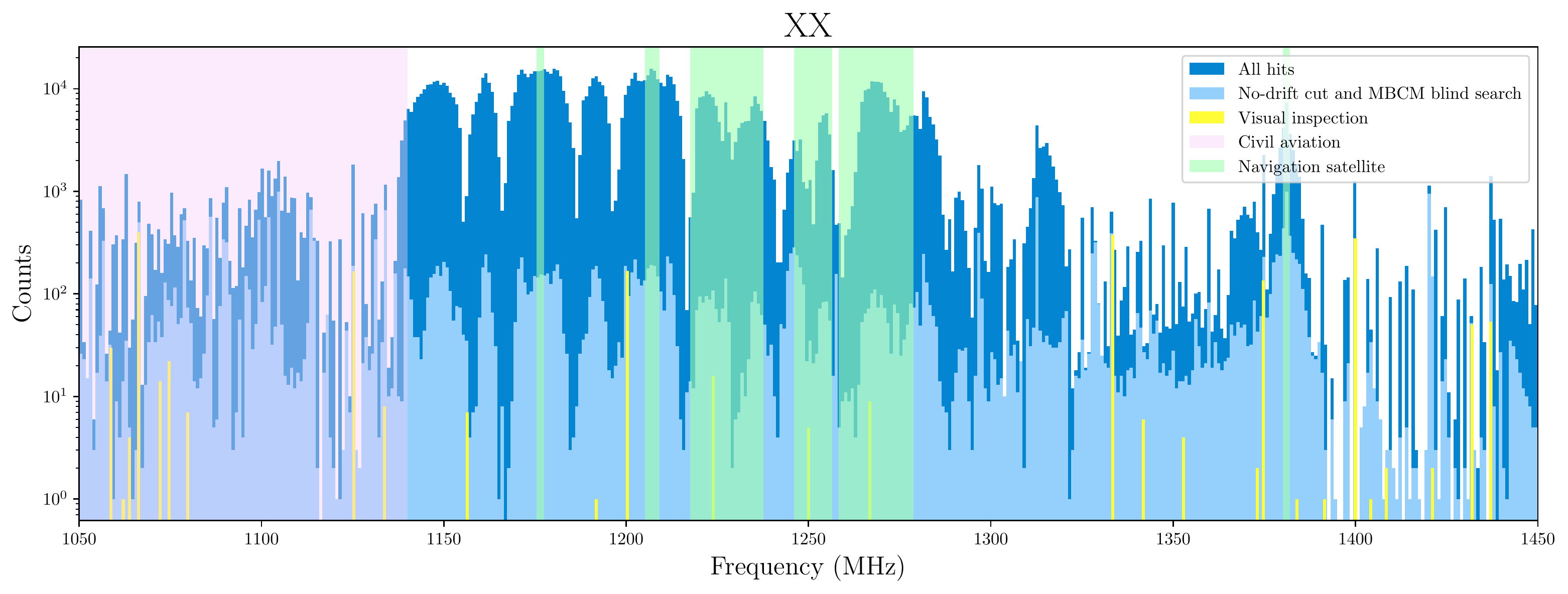}
	}
	
	\subfigure[]{
		\includegraphics[width=0.47\linewidth]{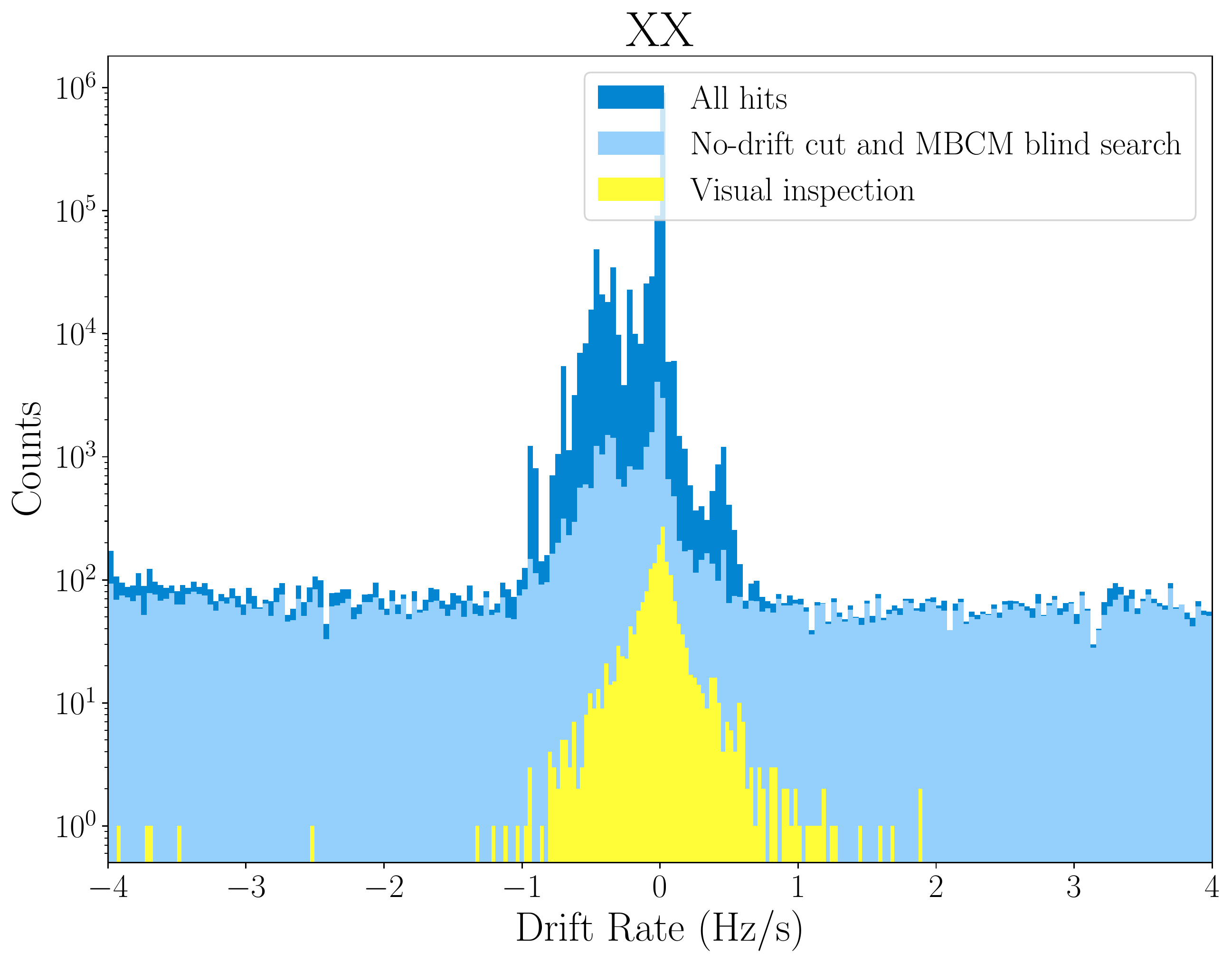}
	}\subfigure[]{
		\includegraphics[width=0.47\linewidth]{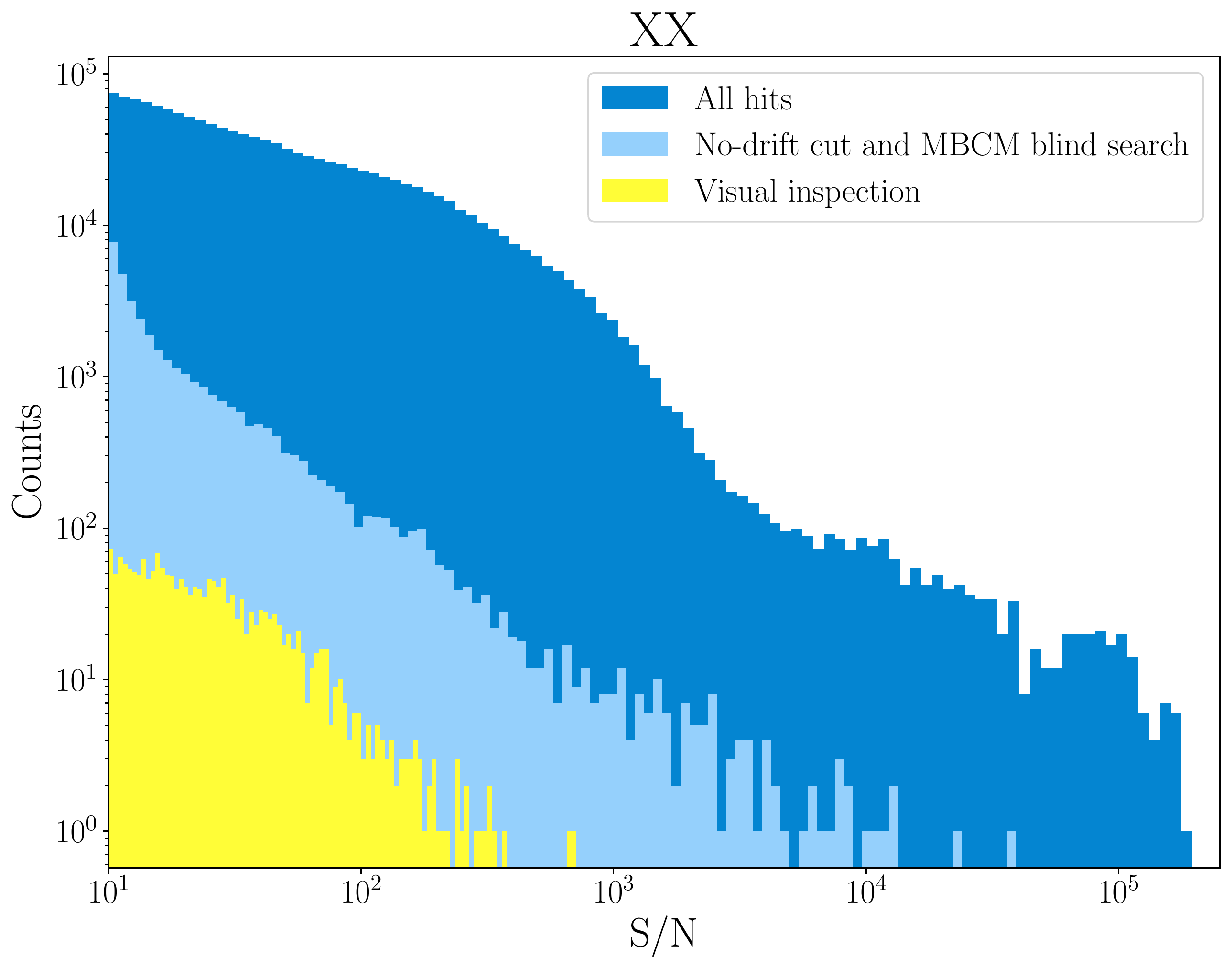}
	}
	\caption{The distributions of frequency, drift rate, and S/N in $XX$ polarization. Frequency bands of catalogued interference sources are displayed on the frequency panel. The dark blue bars show the distributions of the $hits$ detected by all 19 beams. The light blue bars show the distributions of the $events$ removed the zero drift rate signals and detected by the MBCM blind search mode. The yellow bars show the $events$ passed by manual visual inspection.}
	\label{fig:statistcs_XX}
\end{figure*}

\begin{figure*}[!htp]
	
	\centering
	\subfigure[]{
		\includegraphics[width=0.98\linewidth]{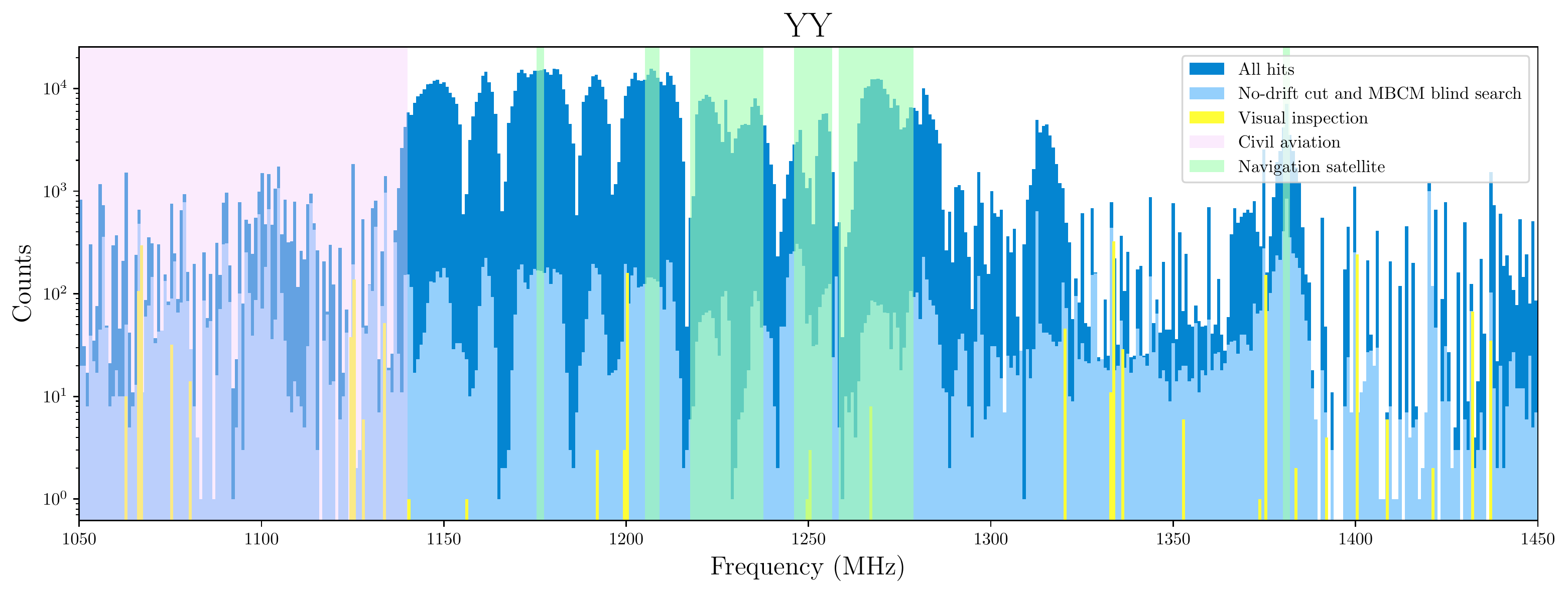}
	}
	
	\subfigure[]{
		\includegraphics[width=0.47\linewidth]{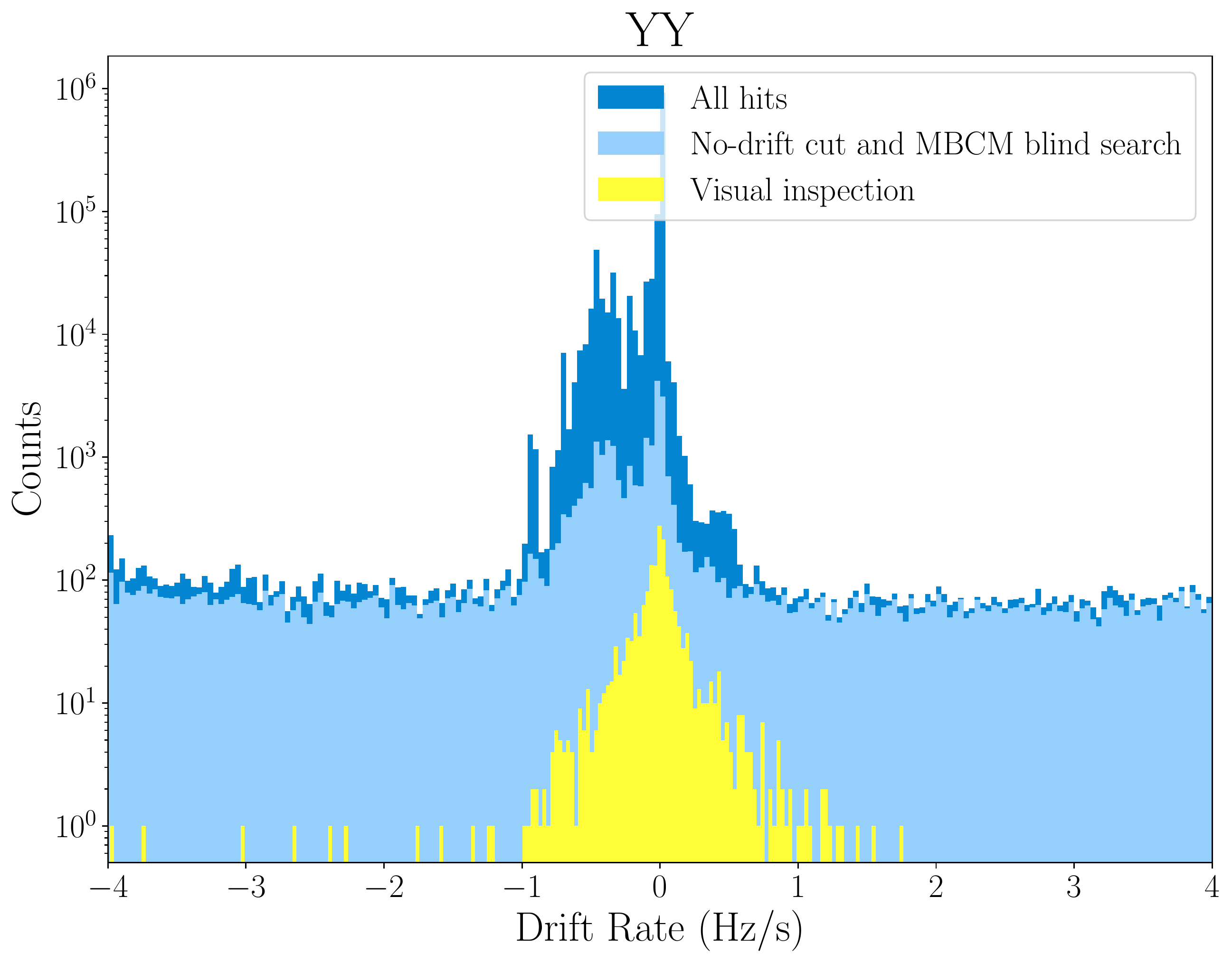}
	}\subfigure[]{
		\includegraphics[width=0.47\linewidth]{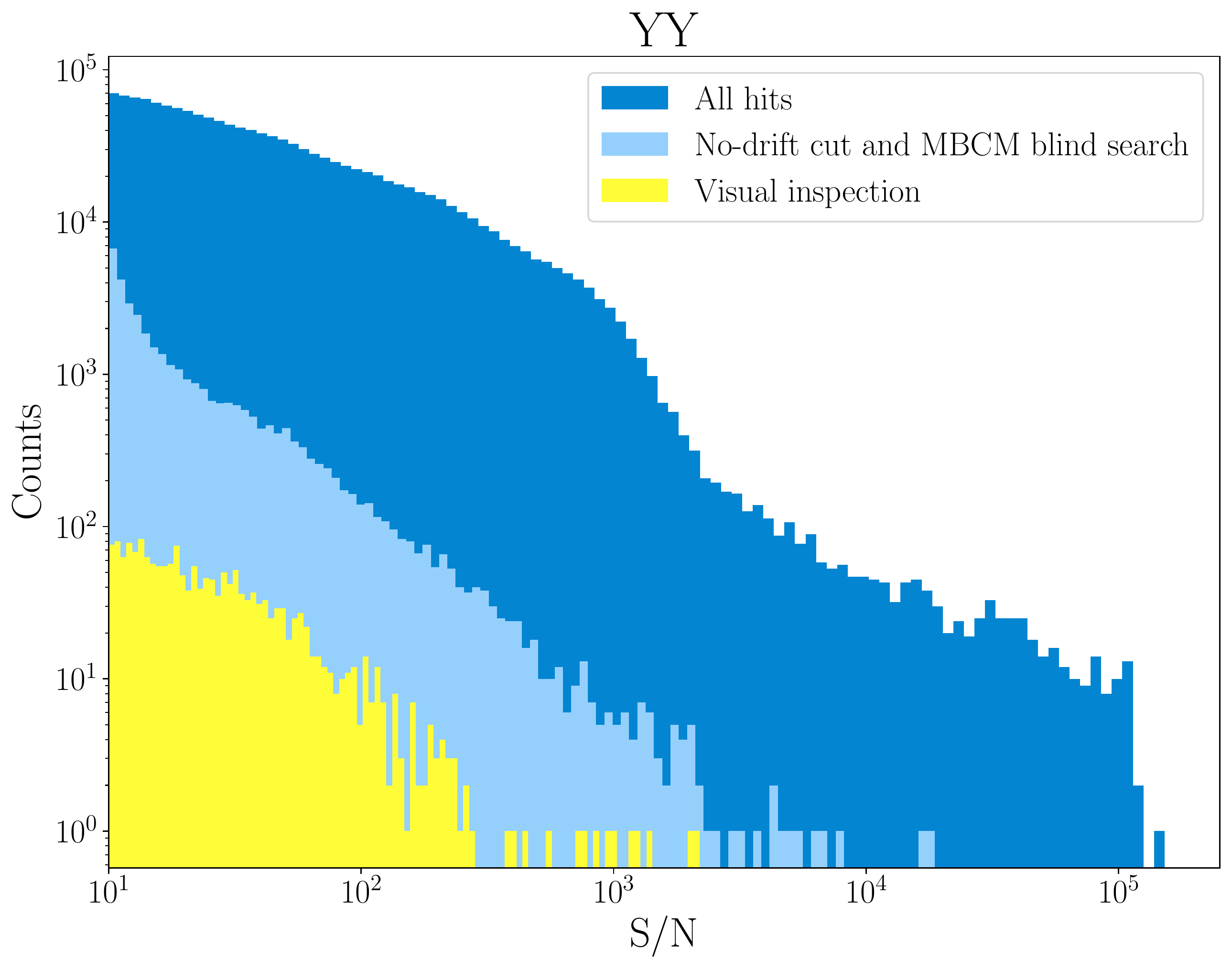}
	}
	\caption{The distributions of frequency, drift rate, and S/N in $YY$ polarization.}
	\label{fig:statistcs_YY}
\end{figure*}

\section{Results} \label{sec:results}

We apply the search pipeline to each of the 20 minute observations from the 33 exoplanet systems (except HD-111998, which lasts $\sim$ 4 minutes). Using an SNR detection threshold of 10 and a maximum drift rate of $\pm$ 4 Hz/s, we search the intensity data of two orthogonal linear polarization directions ($XX$ and $YY$) separately for narrowband signals across $1.05-1.45$ GHz (L-band) and result in about 1.3 millions $hits$. The majority of $hits$ are rejected, and the rejection mechanism is as follows:

\begin{enumerate}
	\item We remove any $hit$ with zero drift rate in the topocentric frame. Those signals most likely correspond to ground-based RFI;
	\item  According to the MBCM blind search mode scheme in Section \ref{subsec:strategy}, we only select signals that follow the allowed beam coverage arrangements;
	\item We re-examine all signals by visual inspection of the dynamic time–frequency spectra (waterfall plots) because most of the signals selected by the program are false positives.
\end{enumerate}

\subsection{Signal Statistics and Identification} \label{subsec:statistics}

We find 1,309,503 $hits$ from $XX$ and 1,324,198 $hits$ from $YY$ in all 19 beams. After removing the no-drift $hits$, we select 34,542 $events$ of $XX$ (89.24\%, 9.51\%, 0.85\% and 0.40\% for 1, 2, 3, and 4 adjacent beams, respectively) and 34,838 $events$ of $YY$ (90.77\%, 8.15\%, 0.71\% and 0.37\% for 1, 2, 3, and 4 adjacent beams, respectively) for MBCM blind search mode. It indicates that the detected signals decrease with increasing complexity of the beam coverage arrangements. As mentioned in the rejection mechanism, visual inspections on the waterfall plots are required for the selected $events$. We find most $events$ are obvious false positives and can be deleted directly. These false positive $events$ have two forms: (1) there is not any signal present in a beam, but TurboSETI sees it as a $hit$; and (2) there is an obvious signal present in a beam, but TurboSETI doesn't determine it a $hit$. For the first case, we can de-drift and time-integrate the spectrum for this beam to prove that there is no signal. There are two potential causes for the second case: (1) the signals are weak and don't reach the S/N threshold in some beams; and (2) there are $hits$ in some beams, but the fitted frequencies are slightly different from other beams. The difference in frequencies exceeds the RFI exclusion frequency range. After removing these false positive $events$, we are left with 1,841 $events$ of $XX$, and 1,804 $events$ of $YY$. We combine the remaining $events$ of $XX$ and $YY$. $Events$ generated by the same signal in $XX$ and $YY$ are considered the same $event$. Therefore, a total of 2055 distinct $events$ passed our visual inspection.

\begin{figure*}
	\centering
	\includegraphics[width=0.85\linewidth]{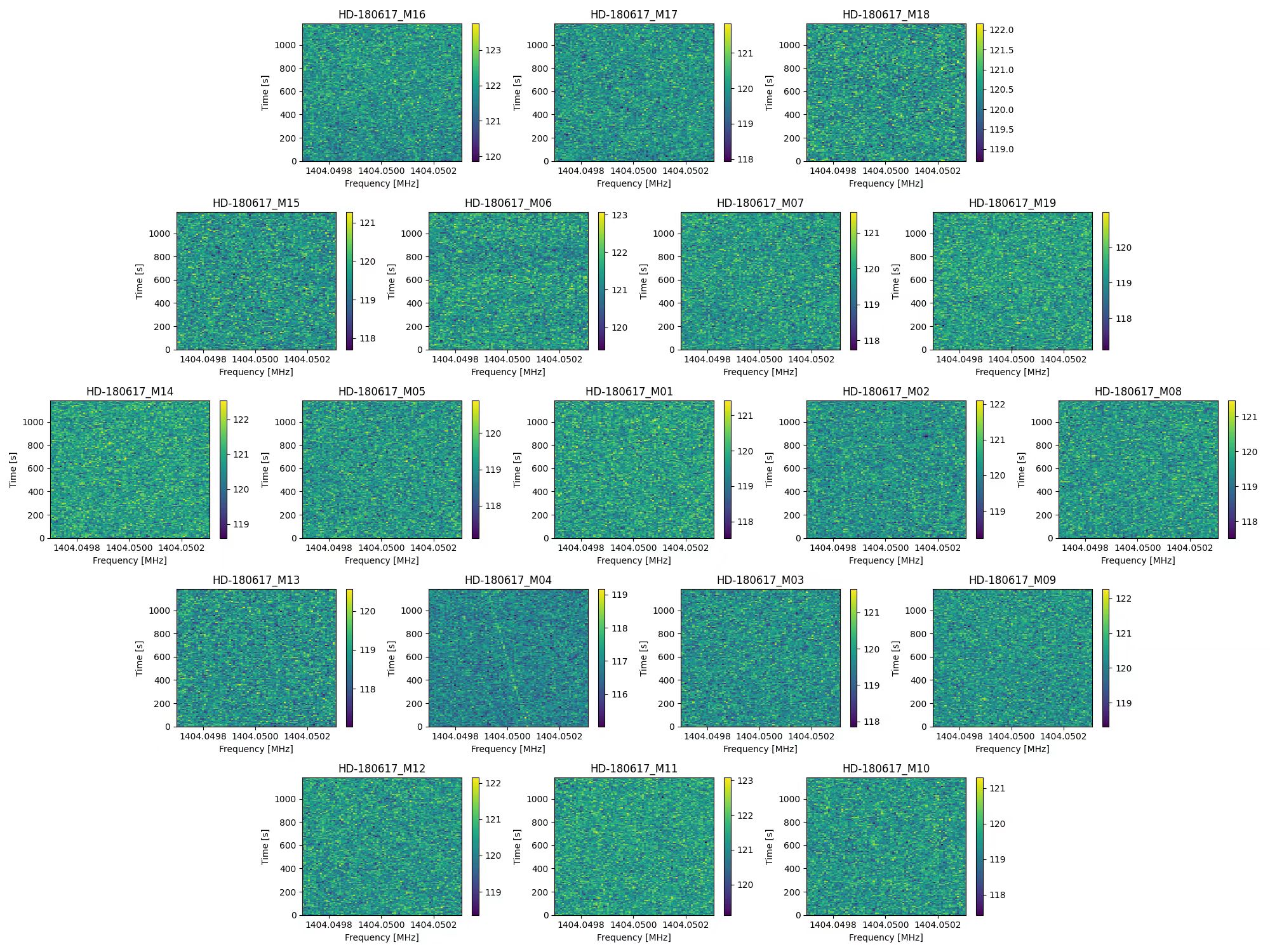}
	\caption{An $event$ detected at 1404.050 MHz toward HD-180617 (NBS 210421). This figure shows $XX$ polarization of all 19 beams.}
	\label{fig:HD-180617_XX_19}
\end{figure*}

The distributions of all $hits$ detected in 19 beams and the $events$ detected by the MBCM blind search mode and passed by visual inspection are shown in Figures \ref{fig:statistcs_XX} and \ref{fig:statistcs_YY}. Each figure illustrates one polarization direction. In these two figures, if one signal is detected by multiple beams, it's parameters are represented by the average values of those beams. As can be seen from Figures \ref{fig:statistcs_XX} and \ref{fig:statistcs_YY}, there is no indication of significant difference between the distributions of $XX$ and $YY$ polarization, which reveals the fact that most $hits$ and $events$ have no preference in term of linear polarization $X$ and $Y$. According to the environmental monitoring of RFI at the FAST site, there are two primary sources of RFI in the frequency band $1.05-1.45$ GHz: civil aviation and navigation satellites \citep{2021RAA....21...18W}. There are only a few $hits$ in the frequency band ($1030 - 1140$ MHz) of civil aviation. However, the number of $hits$ located in the frequency band of the navigation satellite (1176.45 $\pm$ 1.023 MHz, 1207.14 $\pm$ 2.046 MHz, 1227.6 $\pm$ 10 MHz, 1246.0 $\pm$ 1256.5 MHz, 1268.52 $\pm$ 10.23 MHz, 1381.05 $\pm$ 1.023 MHz) shows a higher proportion than for civil aviation, indicating that navigation satellites are potential candidates of primary RFI sources.

The frequency distributions of these 2055 $events$ are shown in Figures \ref{fig:statistcs_XX} and \ref{fig:statistcs_YY}. We find that the frequencies of most events are mostly concentrated around few specific values. For example: 404 $events$ are found at 1066.66072 $\pm$ 0.02702 MHz; 376 $events$ are found at 1333.34161 $\pm$ 0.01878 MHz; 349 $events$ are found at 1400.01084 $\pm$ 0.02630 MHz; and 176 $events$ are found at 1124.98923 $\pm$ 0.01076 MHz. These frequencies can all be linearly combined by the nominal frequencies (33.3333 MHz and 125.00 MHz) of the crystal oscillator on the Roach2 FPGA board, so these $events$ are essentially crystal harmonics. The interferential signals caused by instruments in the backend are known as the instrument RFI \citep{Tao_2022}. The mechanisms causing RFI can be extremely complex. To put it simply, since Roach2 requires crystal oscillators of various frequencies to provide the clock signal, instrument RFI is mainly generated inside the Roach2 FPGA board. Crystal oscillators could potentially introduce instrument RFIs with frequency at linear combinations of the oscillators' nominal frequencies \footnote{\url{https://github.com/ska-sa/roach2_hardware/blob/master/release/rev2/A/BOM/ROACH-2_REV2_BOM.csv}}. In reality, the frequency of a crystal oscillator is unstable due to temperature, voltage, device ageing, and other factors. It may vary in the range of hundreds to thousands of hertz. Since every two beams in the backend share the same Roach2 FPGA board (except for Beam 19) (See Section \ref{subsec:hardware}), the clock oscillator in the same Roach2 can affect the data of two adjacent beams simultaneously.

\subsection{Most Notable Events} \label{subsec:most special}

After the MBCM blind search, visual inspection and removal of instrument RFI, only two $events$ remained. The first $event$ is detected at 1140.604 MHz toward Kepler-438 and its drift rate is about -0.0678 Hz/s, and it is only seen by Beam 1 and not by any other beam. We find that the the signal is much stronger in the $YY$ polarization ($\mathrm{S} / \mathrm{N} \sim 22.64$) than in $XX$ polarization ($\mathrm{S} / \mathrm{N}$ is $\sim$ 9.60). This $event$ is also detected by MBCM targeted search mode (hereafter NBS 210629, where NBS means narrowband signal, and 210629 is the date we detected it). As stated in \citet{Tao_2022}, NBS 210629 is excluded from a ETI signal by certain polarization characteristics.

\begin{figure*}[!htp]
	\centering
	\includegraphics[width=0.98\linewidth]{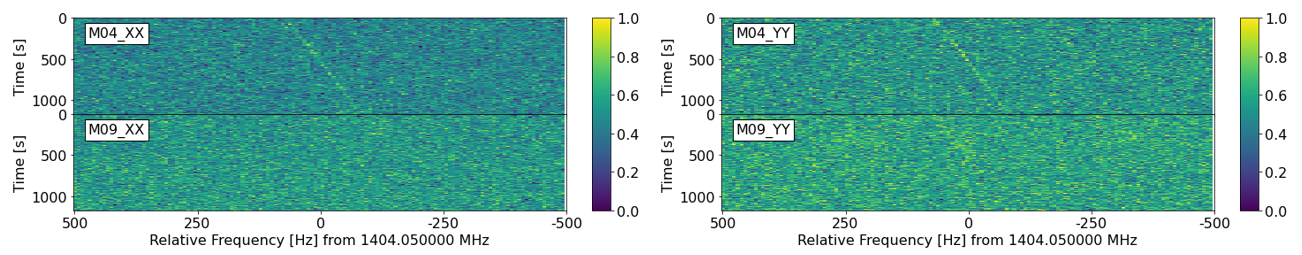}
	\caption{The waterfall plots of Beams 4 and 9 for HD-180617 NBS 210421 in $XX$ and $YY$ by the de-drifting algorithm.}
	\label{fig:1404}
\end{figure*}

The second $event$ is detected at 1404.050 MHz toward HD-180617 and its drift rate is about -0.096 Hz/s (hereafter NBS 210421). Its frequency cannot be obtained by a linear combination of the nominal frequencies of crystal oscillators, and its drift rate is within reach of a transmitter moving with an exoplanet. But it is not present in the central beam. It is only present in Beam 4. The coordinates of Beam 4 for HD-180617 are RA = 19:16:43.24, DEC = +05:04:47.9. Two stars are archived within three arc min, ATO J289.1824+05.1208 and TYC 472-734-1. By visual inspection, NBS 210421 is found in $XX$ polarization (Figure \ref{fig:HD-180617_XX_19}). This $event$ is only detected by the MBCM blind search mode. The fact that NBS 210421 is only detected by MBCM blind search mode displays the advantage on effectiveness and comprehensiveness of our new mode. To analyze it further, we use the de-drifting algorithm for all 19 beams of $XX$ and $YY$ separately. The $\mathrm{S} / \mathrm{N}$s for Beam 4 of $XX$ and $YY$ are not much different ($\mathrm{S} / \mathrm{N}$ $\sim$ 5), and their drift rates are roughly at the same level. But de-drift algorithm also find that Beam 9 (spaced from Beam 4 at the receiver) of the $YY$ polarization at the same frequency also faintly presents a narrowband signal with a similar drift rate. However, the signal of Beam 9 is completely invisible during our visual inspection. More detailed information about NBS 210421 obtained by the de-drifting algorithm is shown in Figure \ref{fig:1404} and Table \ref{table:special}.

According to Section 4.3 of \citet{Tao_2022}, instrument RFIs from the same source have similar polarization characteristics. Therefore, we visually examine the waterfall plots of all 19 beams at 1404.050 MHz $\pm$ 2 kHz in both $XX$ and $YY$ among the observations toward other targets. Nine targets (HD-197037, HD-111998, Wolf-1061, HD-210277, GJ-3323, HD-30562, GJ-96, HD-82943, HD-69830) are detected to have narrowband signals in each of their Beam 4 and 9 simultaneously. But these signals are almost no-drifting and do not appear related to the chronological order of observation. At this stage, we are still unclear weather there is connection between these signals and NBS 210421.

\begin{table*}[!ht]
	\caption{The detailed information of HD-180617 NBS 210421}
	\centering
	\begin{tabular}{ccccccc}
		\hline \hline Beam No. & Polarization & Frequency & Decimal MJD & Drift Rate & S/N & Visual judgment\\
		\colhead{} & \colhead{} & (MHz)  & \colhead{} & (Hz $s^{-1})$ & \colhead{} & (yes/no) \\
		\hline M04 & $XX$ & $1404.0500$ & $59325.003808$ & $-0.096$ & $5.37$ & y\\
		M04 & $YY$ & $1404.0500$ & $59325.003808$ & $-0.096$ & $4.86$ & y\\
		\hline M09 & $XX$ & -- & -- & -- & -- & n\\
		M09 & $YY$ & $1404.0500$ & $59325.003808$ & $-0.072$ & $4.36$ & n\\
		\hline
	\end{tabular}
	\label{table:special}
\end{table*}

\begin{figure*}[!htp]
	\centering
	\subfigure[]{
		\includegraphics[width=0.98\linewidth]{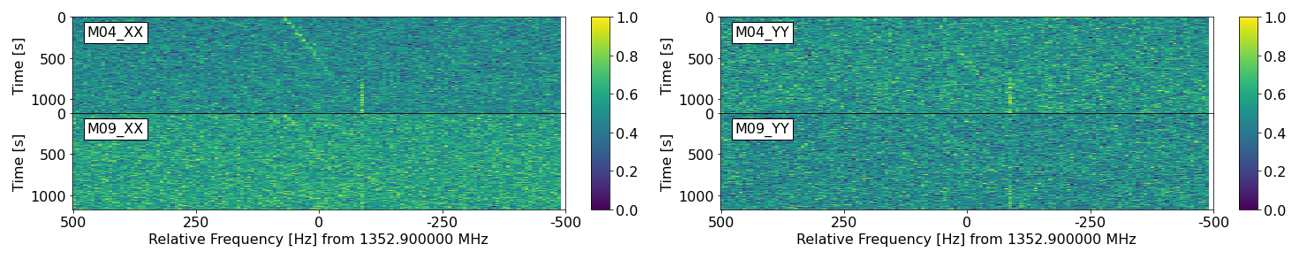}
	}
	
	\subfigure[]{
		\includegraphics[width=0.98\linewidth]{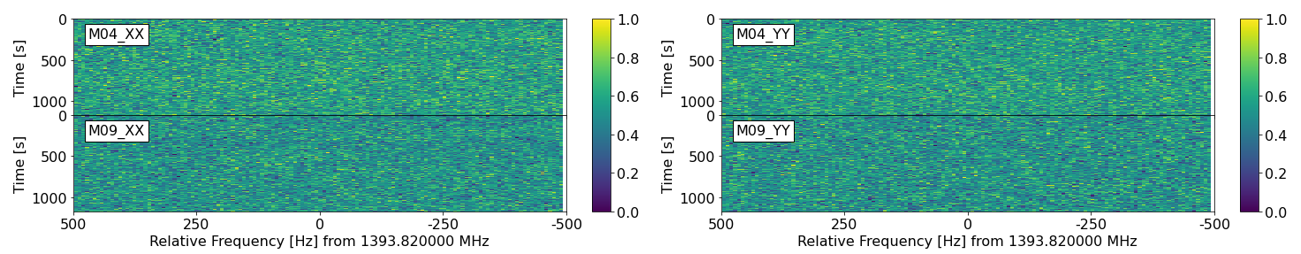}
	}
	
	\subfigure[]{
		\includegraphics[width=0.98\linewidth]{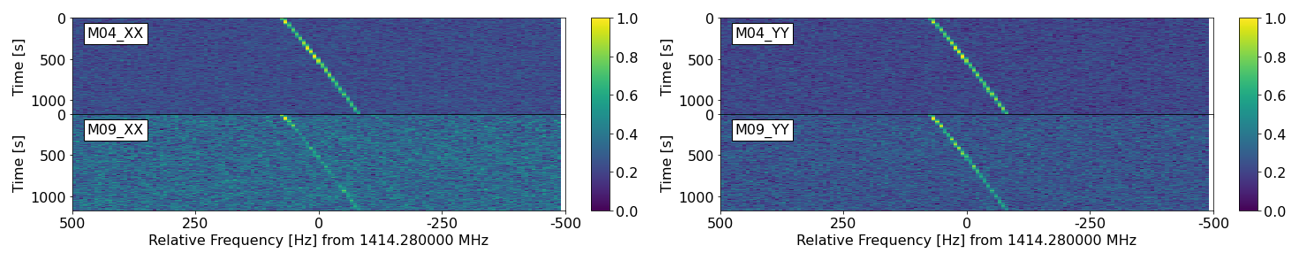}
	}
	
	\subfigure[]{
		\includegraphics[width=0.98\linewidth]{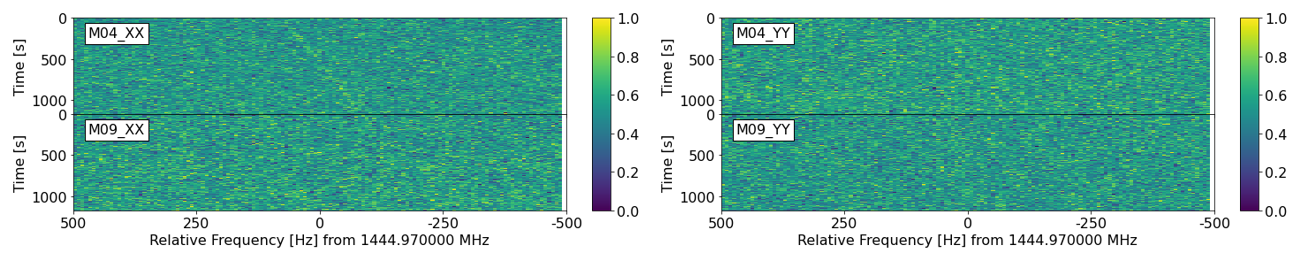}
	}
	\caption{The other four signals of HD-180617 are highly consistent with the characteristics of NBS 210421. The left column is $XX$ polarization, and the right column is $YY$ polarization. The beam coverage, drift rate, and polarization characteristics are consistent with NBS 210421. These frequencies are different from 1404.050MHz by multiples of 10.23MHz.}
	\label{fig:others}
\end{figure*}

Then we re-examine the signals for the same observation toward HD-180617 at other frequencies and find four narrowband drifting signals whose frequencies are around 1352.900 MHz, 1393.820 MHz, 1414.280 MHz, and 1444.970 MHz (Figure \ref{fig:others}). These frequencies are around 1404.050 MHz, and their differences are all multiples of 10.23 MHz. Some of these signals are only present in Beam 4 and Beam 9, and their drift behaviours are very similar to NBS 210421. Most importantly, we find that the $XX$ and $YY$ polarization intensities of all four signals are approximately equal. In a word, the four signals and NBS 210421 show similar features as listed below:
\begin{enumerate}

\item [(1)] similar beam coverage (Beam 4 and Beam 9);
\item [(2)] similar drift rate; 
\item [(3)] similar polarization characteristics (similar intensities in $XX$ and $YY$); 
\item [(4)] the frequencies differ by multiples of 10.23 MHz from 1404.050 MHz (1352.900 MHz, 1393.820 MHz, 1414.280 MHz, and 1444.970 MHz).

\end{enumerate}
Based on the above characteristics, NBS 210421 is almost certain to come from a same instrumental source same as these four signals, but their origin is still unknown.


\section{Conclusions and Discussions} \label{sec:discussion and conclusions}

We conduct searches for narrowband drifting radio signals toward 33 exoplanet systems across $1.05-1.45$ GHz using the FAST 19-beam receiver. With the maximum effective aperture (300 m) and extremely low system temperature ($\text{T}_{\text{sys}}$ between $18-24$ K), the sensitivity of FAST is $\sim 2000\;m^{2}\; K^{-1}$. The minimum equivalent isotropic radiated power ($\text{EIRP}_{\text{min}}$) we can detect reaches $1.48 \times 10^{9}$ W \citep{Tao_2022}, so our targeted SETI observations achieve an unprecedented sensitivity.

In this paper, a new search mode called MBCM blind search mode is proposed in targeted SETI observations and reveals huge advantages. With this new blind search mode, we detect two interesting signals, one of which can only be detected by the blind search mode while the other is detected by both blind and targeted search modes. Furthermore, we exclude the possibility that either signal is due to ETI.

\subsection{Advantages of MBCM Blind Search mode} \label{subsec:blind search}

Compared with the traditional on-off strategy \citep{2013ApJ...767...94S,2017ApJ...849..104E,2019AJ....157..122P,2020AJ....159...86P,2020AJ....160...29S,2021AJ....161..286T,2021NatAs...5.1148S,Gajjar_2021} and the MBCM targeted search mode \citep{Tao_2022}, the MBCM blind search mode shows many significant advantages.

First of all, the blind search mode can comprehensively utilize all 19 beams compared with the MBCM targeted search mode. It can detect signals within $\sim$ $11.6^{\prime}$ of the observation targets. Apart from detecting signals in one beam, it also detects signals between beams. There may be undiscovered exoplanets scattered around the pointed targets, so searching the larger sky near the targets may have new discoveries. According to \citet{Tao_2022}, the MBCM targeted search mode only uses seven beams. MBCM blind search mode works as an extension to the MBCM targeted search mode by adding extra allowed beam coverage arrangements. According to the results of these two researches, NBS 210629 is detected by both search modes, and NBS 210421 is detected only by the MBCM blind search mode. Therefore, the MBCM blind search mode can detect extra signals that could potentially be interesting.

Secondly, beam coverage arrangements can be used to identify instrument RFI. As mentioned in Section \ref{subsec:hardware}, the clock oscillator harmonics appear in two adjacent beams at the same time because the two beams share the same Roach2 FPGA board. Thus, RFI caused by sharing Roach2 FPGA board is usually detected by two adjacent beams.

Lastly, the traditional on-off strategy cannot identify RFI whose duty cycle coincides with the on-off observation period. If RFI happens to appear during the on-observation and disappear during the off-observations, then the on-off strategy will fail to spot it. Since the 19 beams used in MBCM blind search mode observes the sky regions near the targets simultaneously, this coincidence can be avoided.

\begin{figure*}
	\centering
	\includegraphics[width=0.98\linewidth]{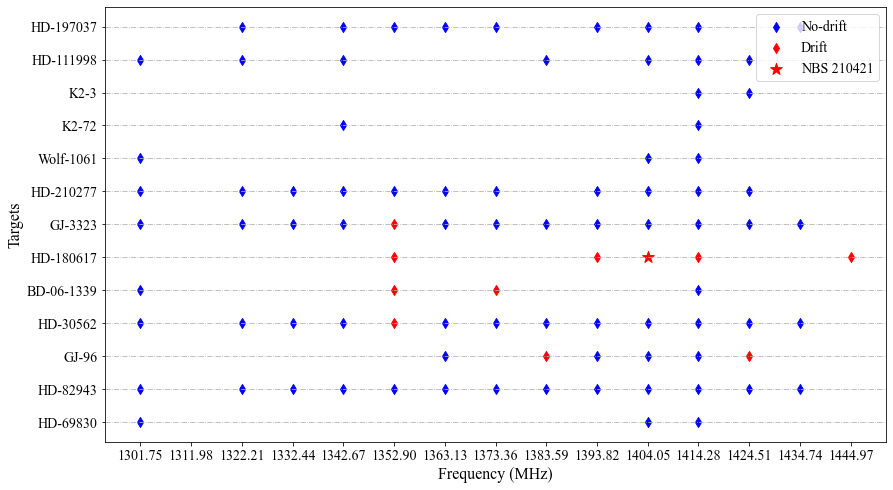}
	\caption{Statistics of the signals present only in Beams 4 and 9 of $XX$ and $YY$. The blue dots represent signals with zero drift rates and the red dots represent signals with non-zero drift rates. The red star represents NBS 210421. In these particular targets, the frequencies of all signals form an evenly spaced regularity, which reveals an unknown instrumental RFI. NBS 210421 and the other four signals of HD-180617 are included in this regularity.}
	\label{fig:Beams 4&9}
\end{figure*}

\subsection{The discussions of polarization analysis}
\label{subsec:polarization}

In the MBCM strategy, apart from comparing multiple beams, we also process the data separately for a single beam in two orthogonal linear polarization directions ($XX$ and $YY$). Polarization characteristics provide strong evidence for the rejection of instrumental RFI.


We assume that an ETI detection would presumably show similar intensities in the spectra of $XX$ and $YY$ polarization in most cases. However, the intensity in one polarization channel ($XX$ or $YY$) can be much stronger than the other if the incoming ETI signal is linearly polarized and aligned precisely with the $X$ or $Y$ polarization of the feed. In addition, the intensities of $XX$ and $YY$ will vary with the parallax angle of the source when the observation duration is extended. Therefore, an intensity difference between $XX$ and $YY$ does not guarantee its origin. We should focus on the variation of polarization intensity over a longer observation time.

Apart from intensity difference and variation over time between $XX$ and $YY$ polarization, the polarization relationship between the particular signal and other signals can serve as great tool for identifying instrumental RFI effectively. \citet{Tao_2022} measured the $XX$ and $YY$ S/Ns of 298 clock oscillator harmonics, concluding that instrumental RFI from the same source has similar polarization characteristics. Polarization analysis includes the following two aspects: firstly, detections in different target observations with roughly the same frequency and similar polarization characteristics, which implies there is a higher chance that they come from the same origin, such as NBS 210629 and eight other signals in \citet{Tao_2022}; secondly, when detections from the same target observation shows correlations in frequency, we can then look at their polarization characteristics. A similar polarization characteristics of detections could suggest that they are from the same origin, such as NBS 210421 and the other four signals in this paper.

However, it should be emphasized that polarization characteristics are not the only criterion for identifying candidate ETI signals. We must combine many other factors to analyze a particular signal, including beam coverage, drift rate, frequency relations, and other characteristics.


In conclusion, polarization characteristics are one of the criteria for identifying the origin of a signal. It improves the efficiency for signal identification. In addition to the MBCM strategy, polarization analysis can also be useful to the on-off strategy as long as polarization directions are recorded separately.

\subsection{The particularity of Beam 4 and Beam 9} \label{subsec:beams 4&9}

Although Beam 4 and Beam 9 are not adjacent on the 19-beam receiver, they both indicate related characteristics between NBS 210421 and the other four signals of HD-180617. Thus, we examine all the signals detected by TurboSETI present only in Beam 4 and Beam 9 and find the following pattern. Excluding the frequency bands of civil aviation and navigation satellites, there are many narrowband signals that only appear in Beam 4 or Beam 9 or both. The frequencies seem to roughly form an arithmetic sequence starting at 1301.750 MHz with a common difference of around 10.23 MHz (most of the frequencies are not from clock oscillator harmonics). The intensities of $XX$ and $YY$ of these signals are approximately the same. Most of these signals are no-drift, and a small number of signals show drifting features. In addition, these signals are always detected from the sky locations near the 13 targets listed in the Y-axis in Figure \ref{fig:Beams 4&9}.

Statistics of the signals present only in Beam 4 and Beam 9 of $XX$ and $YY$ are shown in Figure \ref{fig:Beams 4&9}. As stated in Figure \ref{fig:Beams 4&9}, the frequencies of NBS 210421 and the other four signals of HD-180617 differ by multiples of 10.23 MHz from 1404.050 MHz (1352.900 MHz, 1393.820 MHz, 1414.280 MHz and 1444.970 MHz). By testing the Roach2 board, we find there is a bus signal that never leaves the PowerPC. It causes spurts at certain frequencies (and a few harmonics) and produces signals at these certain frequency bands, but the exact origin is still unknown. Thus, NBS 210421 and the other four signals are most likely to come from instrumental RFI.

To achieve all key science goals, FAST will conduct more SETI observations in the future, including targeted searches and commensal sky surveys. Since FAST is the most sensitive single-dish radio telescope on Earth, SETI with FAST is challenged by more types of weak RFI. We will identify and classify more types of RFI and discover new search methods to effectively remove RFIs. In addition, we will improve the FAST multibeam digital backend to shield the instrument RFI and enhance the SETI performance of FAST.

~\\

We sincerely appreciate the referee’s suggestions, which help us greatly improve our manuscript. We also sincerely thank Wei Hong for the kind and useful discussions. This work was supported by Chinese SKA–the Cradle of Life of the Ministry of Science and Technology of China, the National Science Foundation of China (Grants No.11929301), and the National Key R$\&$D Program of China (2017YFA0402600). Shi-Yu Li is supported by Beijing Postdoctoral Research Foundation. This work is finished on the servers from FAST Data Center in Dezhou University.

\bibliography{arXiv-0225}{}

\begin{thebibliography}{}
\expandafter\ifx\csname natexlab\endcsname\relax\def\natexlab#1{#1}\fi
\providecommand{\url}[1]{\href{#1}{#1}}
\providecommand{\dodoi}[1]{doi:~\href{http://doi.org/#1}{\nolinkurl{#1}}}
\providecommand{\doeprint}[1]{\href{http://ascl.net/#1}{\nolinkurl{http://ascl.net/#1}}}
\providecommand{\doarXiv}[1]{\href{https://arxiv.org/abs/#1}{\nolinkurl{https://arxiv.org/abs/#1}}}

\bibitem[{{Batalha}(2014)}]{2014PNAS..11112647B}
{Batalha}, N.~M. 2014, Proceedings of the National Academy of Science, 111,
  12647, \dodoi{10.1073/pnas.1304196111}

\bibitem[{{Cocconi} \& {Morrison}(1959)}]{1959Natur.184..844C}
{Cocconi}, G., \& {Morrison}, P. 1959, \nat, 184, 844, \dodoi{10.1038/184844a0}

\bibitem[{{Cohen} {et~al.}(1987){Cohen}, {Downs}, {Emerson}, {Grimm}, {Gulkis},
  {Stevens}, \& {Tarter}}]{1987MNRAS.225..491C}
{Cohen}, R.~J., {Downs}, G., {Emerson}, R., {et~al.} 1987, \mnras, 225, 491,
  \dodoi{10.1093/mnras/225.3.491}

\bibitem[{{Drake}(1961)}]{1961PhT....14d..40D}
{Drake}, F.~D. 1961, Physics Today, 14, 40, \dodoi{10.1063/1.3057500}

\bibitem[{{Dressing} \& {Charbonneau}(2013)}]{2013ApJ...767...95D}
{Dressing}, C.~D., \& {Charbonneau}, D. 2013, \apj, 767, 95,
  \dodoi{10.1088/0004-637X/767/1/95}

\bibitem[{{Enriquez} \& {Price}(2019)}]{2019ascl.soft06006E}
{Enriquez}, E., \& {Price}, D. 2019, {turboSETI: Python-based SETI search
  algorithm}, Astrophysics Source Code Library, record ascl:1906.006.
\newblock \doeprint{1906.006}

\bibitem[{{Enriquez} {et~al.}(2017){Enriquez}, {Siemion}, {Foster}, {Gajjar},
  {Hellbourg}, {Hickish}, {Isaacson}, {Price}, {Croft}, {DeBoer}, {Lebofsky},
  {MacMahon}, \& {Werthimer}}]{2017ApJ...849..104E}
{Enriquez}, J.~E., {Siemion}, A., {Foster}, G., {et~al.} 2017, \apj, 849, 104,
  \dodoi{10.3847/1538-4357/aa8d1b}

\bibitem[{Gajjar {et~al.}(2021)Gajjar, Perez, Siemion, Foster, Brzycki,
  Chatterjee, Chen, Cordes, Croft, Czech, DeBoer, DeMarines, Drew, Gowanlock,
  Isaacson, Lacki, Lebofsky, MacMahon, Morrison, Ng, de~Pater, Price, Sheikh,
  Suresh, Webb, \& Worden}]{Gajjar_2021}
Gajjar, V., Perez, K.~I., Siemion, A. P.~V., {et~al.} 2021, The Astronomical
  Journal, 162, 33, \dodoi{10.3847/1538-3881/abfd36}

\bibitem[{{Gray} \& {Mooley}(2017)}]{2017AJ....153..110G}
{Gray}, R.~H., \& {Mooley}, K. 2017, \aj, 153, 110,
  \dodoi{10.3847/1538-3881/153/3/110}

\bibitem[{{Harp} {et~al.}(2020){Harp}, {Gray}, {Richards}, {Shostak}, \&
  {Tarter}}]{2020AJ....160..162H}
{Harp}, G.~R., {Gray}, R.~H., {Richards}, J., {Shostak}, G.~S., \& {Tarter},
  J.~C. 2020, \aj, 160, 162, \dodoi{10.3847/1538-3881/aba58f}

\bibitem[{{Harp} {et~al.}(2016){Harp}, {Richards}, {Tarter}, {Dreher},
  {Jordan}, {Shostak}, {Smolek}, {Kilsdonk}, {Wilcox}, {Wimberly}, {Ross},
  {Barott}, {Ackermann}, \& {Blair}}]{2016AJ....152..181H}
{Harp}, G.~R., {Richards}, J., {Tarter}, J.~C., {et~al.} 2016, \aj, 152, 181,
  \dodoi{10.3847/0004-6256/152/6/181}

\bibitem[{{Harp} {et~al.}(2018){Harp}, {Ackermann}, {Astorga}, {Arbunich},
  {Barrios}, {Hightower}, {Meitzner}, {Barott}, {Nolan}, {Messerschmitt},
  {Vakoch}, {Shostak}, \& {Tarter}}]{2018ApJ...869...66H}
{Harp}, G.~R., {Ackermann}, R.~F., {Astorga}, A., {et~al.} 2018, \apj, 869, 66,
  \dodoi{10.3847/1538-4357/aaeb98}

\bibitem[{{Jiang} {et~al.}(2019){Jiang}, {Yue}, {Gan}, {Yao}, {Li}, {Pan},
  {Sun}, {Yu}, {Liu}, {Tang}, {Qian}, {Lu}, {Yan}, {Peng}, {Zhang}, {Wang},
  {Li}, \& {Li}}]{2019SCPMA..6259502J}
{Jiang}, P., {Yue}, Y., {Gan}, H., {et~al.} 2019, Science China Physics,
  Mechanics, and Astronomy, 62, 959502, \dodoi{10.1007/s11433-018-9376-1}

\bibitem[{{Jiang} {et~al.}(2020){Jiang}, {Tang}, {Hou}, {Liu}, {Kr{\v{c}}o},
  {Qian}, {Sun}, {Ching}, {Liu}, {Duan}, {Yue}, {Gan}, {Yao}, {Li}, {Pan},
  {Yu}, {Liu}, {Li}, {Peng}, {Yan}, \& {FAST
  Collaboration}}]{2020RAA....20...64J}
{Jiang}, P., {Tang}, N.-Y., {Hou}, L.-G., {et~al.} 2020, Research in Astronomy
  and Astrophysics, 20, 064, \dodoi{10.1088/1674-4527/20/5/64}

\bibitem[{{Kaltenegger} \& {Faherty}(2021)}]{2021Natur.594..505K}
{Kaltenegger}, L., \& {Faherty}, J.~K. 2021, \nat, 594, 505,
  \dodoi{10.1038/s41586-021-03596-y}

\bibitem[{Kasting {et~al.}(1993)Kasting, Whitmire, \&
  Reynolds}]{KASTING1993108}
Kasting, J.~F., Whitmire, D.~P., \& Reynolds, R.~T. 1993, Icarus, 101, 108,
  \dodoi{https://doi.org/10.1006/icar.1993.1010}

\bibitem[{Keane {et~al.}(2017)Keane, Barr, Jameson, Morello, Caleb, Bhandari,
  Petroff, Possenti, Burgay, Tiburzi, Bailes, Bhat, Burke-Spolaor, Eatough,
  Flynn, Jankowski, Johnston, Kramer, Levin, Ng, van Straten, \&
  Krishnan}]{10.1093/mnras/stx2126}
Keane, E.~F., Barr, E., Jameson, A., {et~al.} 2017, Monthly Notices of the
  Royal Astronomical Society, 473, 116, \dodoi{10.1093/mnras/stx2126}

\bibitem[{Kopparapu {et~al.}(2013)Kopparapu, Ramirez, Kasting, Eymet, Robinson,
  Mahadevan, Terrien, Domagal-Goldman, Meadows, \& Deshpande}]{Kopparapu_2013}
Kopparapu, R.~K., Ramirez, R., Kasting, J.~F., {et~al.} 2013, The Astrophysical
  Journal, 765, 131, \dodoi{10.1088/0004-637x/765/2/131}

\bibitem[{{Li} \& {Pan}(2016)}]{2016RaSc...51.1060L}
{Li}, D., \& {Pan}, Z. 2016, Radio Science, 51, 1060,
  \dodoi{10.1002/2015RS005877}

\bibitem[{{Li} {et~al.}(2020){Li}, {Gajjar}, {Wang}, {Siemion}, {Zhang},
  {Zhang}, {Yue}, {Zhu}, {Jin}, {Li}, {Berger}, {Brzycki}, {Cobb}, {Croft},
  {Czech}, {DeBoer}, {DeMarines}, {Drew}, {Emilio Enriquez}, {Gizani},
  {Korpela}, {Isaacson}, {Lebofsky}, {Lacki}, {MacMahon}, {Nanez}, {Niu},
  {Pei}, {Price}, {Werthimer}, {Worden}, {Gerry Zhang}, {Zhang}, \& {FAST
  Collaboration}}]{2020RAA....20...78L}
{Li}, D., {Gajjar}, V., {Wang}, P., {et~al.} 2020, Research in Astronomy and
  Astrophysics, 20, 078, \dodoi{10.1088/1674-4527/20/5/78}

\bibitem[{{Li} {et~al.}(2022){Li}, {Zhao}, {Tao}, {Zhang}, \&
  {Xiao-Hui}}]{2022ApJ...938....1L}
{Li}, J.-K., {Zhao}, H.-C., {Tao}, Z.-Z., {Zhang}, T.-J., \& {Xiao-Hui}, S.
  2022, \apj, 938, 1, \dodoi{10.3847/1538-4357/ac90bd}

\bibitem[{{MacMahon} {et~al.}(2018){MacMahon}, {Price}, {Lebofsky}, {Siemion},
  {Croft}, {DeBoer}, {Enriquez}, {Gajjar}, {Hellbourg}, {Isaacson},
  {Werthimer}, {Abdurashidova}, {Bloss}, {Brandt}, {Creager}, {Ford}, {Lynch},
  {Maddalena}, {McCullough}, {Ray}, {Whitehead}, \&
  {Woody}}]{2018PASP..130d4502M}
{MacMahon}, D. H.~E., {Price}, D.~C., {Lebofsky}, M., {et~al.} 2018, \pasp,
  130, 044502, \dodoi{10.1088/1538-3873/aa80d2}

\bibitem[{{Margot} {et~al.}(2021){Margot}, {Pinchuk}, {Geil}, {Alexander},
  {Arora}, {Biswas}, {Cebreros}, {Desai}, {Duclos}, {Dunne}, {Lin Fu}, {Goel},
  {Gonzales}, {Gonzalez}, {Jain}, {Lam}, {Lewis}, {Lewis}, {Li}, {MacDougall},
  {Makarem}, {Manan}, {Molina}, {Nagib}, {Neville}, {O'Toole}, {Rockwell},
  {Rokushima}, {Romanek}, {Schmidgall}, {Seth}, {Shah}, {Shimane}, {Singhal},
  {Tokadjian}, {Villafana}, {Wang}, {Yun}, {Zhu}, \&
  {Lynch}}]{2021AJ....161...55M}
{Margot}, J.-L., {Pinchuk}, P., {Geil}, R., {et~al.} 2021, \aj, 161, 55,
  \dodoi{10.3847/1538-3881/abcc77}

\bibitem[{{Nan}(2006)}]{2006ScChG..49..129N}
{Nan}, R. 2006, Science in China: Physics, Mechanics and Astronomy, 49, 129,
  \dodoi{10.1007/s11433-006-0129-9}

\bibitem[{{Nan} {et~al.}(2000){Nan}, {Peng}, {Zhu}, {Zhu}, {Su}, \&
  {Qui}}]{2000ASPC..213..523N}
{Nan}, R., {Peng}, B., {Zhu}, W., {et~al.} 2000, in Astronomical Society of the
  Pacific Conference Series, Vol. 213, Bioastronomy 99, ed. G.~{Lemarchand} \&
  K.~{Meech}, 523

\bibitem[{{Nan} {et~al.}(2011){Nan}, {Li}, {Jin}, {Wang}, {Zhu}, {Zhu},
  {Zhang}, {Yue}, \& {Qian}}]{2011IJMPD..20..989N}
{Nan}, R., {Li}, D., {Jin}, C., {et~al.} 2011, International Journal of Modern
  Physics D, 20, 989, \dodoi{10.1142/S0218271811019335}

\bibitem[{{Pei} {et~al.}(2019){Pei}, {Li}, {Li}, \&
  {Niu}}]{2019SSPMA..49i9508P}
{Pei}, X., {Li}, J., {Li}, S., \& {Niu}, C. 2019, Scientia Sinica Physica,
  Mechanica \& Astronomica, 49, 099508, \dodoi{10.1360/SSPMA2018-00418}

\bibitem[{{Petigura} {et~al.}(2013){Petigura}, {Howard}, \&
  {Marcy}}]{2013PNAS..11019273P}
{Petigura}, E.~A., {Howard}, A.~W., \& {Marcy}, G.~W. 2013, Proceedings of the
  National Academy of Science, 110, 19273, \dodoi{10.1073/pnas.1319909110}

\bibitem[{{Pinchuk} {et~al.}(2019){Pinchuk}, {Margot}, {Greenberg}, {Ayalde},
  {Bloxham}, {Boddu}, {Gerardo Chinchilla-Garcia}, {Cliffe}, {Gallagher},
  {Hart}, {Hesford}, {Mizrahi}, {Pike}, {Rodger}, {Sayki}, {Schneck}, {Tan},
  {{\textquotedblleft}Yolanda{\textquotedblright} Xiao}, \&
  {Lynch}}]{2019AJ....157..122P}
{Pinchuk}, P., {Margot}, J.-L., {Greenberg}, A.~H., {et~al.} 2019, \aj, 157,
  122, \dodoi{10.3847/1538-3881/ab0105}

\bibitem[{Price {et~al.}(2019)Price, Enriquez, Chen, \& Siebert}]{Price2019}
Price, D.~C., Enriquez, J.~E., Chen, Y., \& Siebert, M. 2019, Journal of Open
  Source Software, 4, 1554, \dodoi{10.21105/joss.01554}

\bibitem[{{Price} {et~al.}(2020){Price}, {Enriquez}, {Brzycki}, {Croft},
  {Czech}, {DeBoer}, {DeMarines}, {Foster}, {Gajjar}, {Gizani}, {Hellbourg},
  {Isaacson}, {Lacki}, {Lebofsky}, {MacMahon}, {Pater}, {Siemion}, {Werthimer},
  {Green}, {Kaczmarek}, {Maddalena}, {Mader}, {Drew}, \&
  {Worden}}]{2020AJ....159...86P}
{Price}, D.~C., {Enriquez}, J.~E., {Brzycki}, B., {et~al.} 2020, \aj, 159, 86,
  \dodoi{10.3847/1538-3881/ab65f1}

\bibitem[{{Rodler} \& {L{\'o}pez-Morales}(2014)}]{2014ApJ...781...54R}
{Rodler}, F., \& {L{\'o}pez-Morales}, M. 2014, \apj, 781, 54,
  \dodoi{10.1088/0004-637X/781/1/54}

\bibitem[{{Roth} {et~al.}(2014){Roth}, {Saur}, {Retherford}, {Strobel},
  {Feldman}, {McGrath}, \& {Nimmo}}]{2014Sci...343..171R}
{Roth}, L., {Saur}, J., {Retherford}, K.~D., {et~al.} 2014, Science, 343, 171,
  \dodoi{10.1126/science.1247051}

\bibitem[{{Schwieterman} {et~al.}(2016){Schwieterman}, {Meadows},
  {Domagal-Goldman}, {Deming}, {Arney}, {Luger}, {Harman}, {Misra}, \&
  {Barnes}}]{2016ApJ...819L..13S}
{Schwieterman}, E.~W., {Meadows}, V.~S., {Domagal-Goldman}, S.~D., {et~al.}
  2016, \apjl, 819, L13, \dodoi{10.3847/2041-8205/819/1/L13}

\bibitem[{{Seager}(2014)}]{2014PNAS..11112634S}
{Seager}, S. 2014, Proceedings of the National Academy of Science, 111, 12634,
  \dodoi{10.1073/pnas.1304213111}

\bibitem[{{Segura} {et~al.}(2005){Segura}, {Kasting}, {Meadows}, {Cohen},
  {Scalo}, {Crisp}, {Butler}, \& {Tinetti}}]{2005AsBio...5..706S}
{Segura}, A., {Kasting}, J.~F., {Meadows}, V., {et~al.} 2005, Astrobiology, 5,
  706, \dodoi{10.1089/ast.2005.5.706}

\bibitem[{{Sheikh} {et~al.}(2020){Sheikh}, {Siemion}, {Enriquez}, {Price},
  {Isaacson}, {Lebofsky}, {Gajjar}, \& {Kalas}}]{2020AJ....160...29S}
{Sheikh}, S.~Z., {Siemion}, A., {Enriquez}, J.~E., {et~al.} 2020, \aj, 160, 29,
  \dodoi{10.3847/1538-3881/ab9361}

\bibitem[{{Siemion} {et~al.}(2013){Siemion}, {Demorest}, {Korpela},
  {Maddalena}, {Werthimer}, {Cobb}, {Howard}, {Langston}, {Lebofsky}, {Marcy},
  \& {Tarter}}]{2013ApJ...767...94S}
{Siemion}, A. P.~V., {Demorest}, P., {Korpela}, E., {et~al.} 2013, \apj, 767,
  94, \dodoi{10.1088/0004-637X/767/1/94}

\bibitem[{Siemion {et~al.}(2013)Siemion, Demorest, Korpela, Maddalena,
  Werthimer, Cobb, Howard, Langston, Lebofsky, Marcy, \& Tarter}]{Siemion_2013}
Siemion, A. P.~V., Demorest, P., Korpela, E., {et~al.} 2013, The Astrophysical
  Journal, 767, 94, \dodoi{10.1088/0004-637x/767/1/94}

\bibitem[{{Smith} {et~al.}(2021){Smith}, {Price}, {Sheikh}, {Czech}, {Croft},
  {DeBoer}, {Gajjar}, {Isaacson}, {Lacki}, {Lebofsky}, {MacMahon}, {Ng},
  {Perez}, {Siemion}, {Webb}, {Drew}, {Worden}, \& {Zic}}]{2021NatAs...5.1148S}
{Smith}, S., {Price}, D.~C., {Sheikh}, S.~Z., {et~al.} 2021, Nature Astronomy,
  5, 1148, \dodoi{10.1038/s41550-021-01479-w}

\bibitem[{Tao {et~al.}(2022)Tao, Zhao, Zhang, Gajjar, Zhu, Yue, Zhang, Liu, Li,
  Zhang, Liu, Wang, Duan, Qian, Jin, Li, Siemion, Jiang, Werthimer, Cobb,
  Korpela, \& Anderson}]{Tao_2022}
Tao, Z.-Z., Zhao, H.-C., Zhang, T.-J., {et~al.} 2022, The Astronomical Journal,
  164, 160, \dodoi{10.3847/1538-3881/ac8bd5}

\bibitem[{{Tarter}(2001)}]{2001ARA&A..39..511T}
{Tarter}, J. 2001, \araa, 39, 511, \dodoi{10.1146/annurev.astro.39.1.511}

\bibitem[{{Tarter} {et~al.}(1980){Tarter}, {Cuzzi}, {Black}, \&
  {Clark}}]{1980Icar...42..136T}
{Tarter}, J., {Cuzzi}, J., {Black}, D., \& {Clark}, T. 1980, \icarus, 42, 136,
  \dodoi{10.1016/0019-1035(80)90251-1}

\bibitem[{{Tingay} {et~al.}(2016){Tingay}, {Tremblay}, {Walsh}, \&
  {Urquhart}}]{2016ApJ...827L..22T}
{Tingay}, S.~J., {Tremblay}, C., {Walsh}, A., \& {Urquhart}, R. 2016, \apjl,
  827, L22, \dodoi{10.3847/2041-8205/827/2/L22}

\bibitem[{{Tingay} {et~al.}(2018){Tingay}, {Tremblay}, \&
  {Croft}}]{2018ApJ...856...31T}
{Tingay}, S.~J., {Tremblay}, C.~D., \& {Croft}, S. 2018, \apj, 856, 31,
  \dodoi{10.3847/1538-4357/aab363}

\bibitem[{{Traas} {et~al.}(2021){Traas}, {Croft}, {Gajjar}, {Isaacson},
  {Lebofsky}, {MacMahon}, {Perez}, {Price}, {Sheikh}, {Siemion}, {Smith},
  {Drew}, \& {Worden}}]{2021AJ....161..286T}
{Traas}, R., {Croft}, S., {Gajjar}, V., {et~al.} 2021, \aj, 161, 286,
  \dodoi{10.3847/1538-3881/abf649}

\bibitem[{{Tremblay} \& {Tingay}(2020)}]{2020PASA...37...35T}
{Tremblay}, C.~D., \& {Tingay}, S.~J. 2020, \pasa, 37, e035,
  \dodoi{10.1017/pasa.2020.27}

\bibitem[{{Wang} {et~al.}(2021){Wang}, {Zhang}, {Hu}, {Huang}, {Zhu}, {Zhi},
  {Zhang}, {Fan}, \& {Yang}}]{2021RAA....21...18W}
{Wang}, Y., {Zhang}, H.-Y., {Hu}, H., {et~al.} 2021, Research in Astronomy and
  Astrophysics, 21, 018, \dodoi{10.1088/1674-4527/21/1/18}

\bibitem[{{Webster} {et~al.}(2015){Webster}, {Mahaffy}, {Atreya}, {Flesch},
  {Mischna}, {Meslin}, {Farley}, {Conrad}, {Christensen}, {Pavlov},
  {Mart{\'\i}n-Torres}, {Zorzano}, {McConnochie}, {Owen}, {Eigenbrode},
  {Glavin}, {Steele}, {Malespin}, {Archer}, {Sutter}, {Coll}, {Freissinet},
  {McKay}, {Moores}, {Schwenzer}, {Bridges}, {Navarro-Gonzalez}, {Gellert},
  {Lemmon}, {MSL Science Team}, {Abbey}, {Achilles}, {Agard}, {Alexandre Alves
  Verdasca}, {Anderson}, {Anderson}, {Anderson}, {Appel}, {Archer}, {Arevalo},
  {Armiens-Aparicio}, {Arvidson}, {Atlaskin}, {Atreya}, {Azeez}, {Baker},
  {Baker}, {Balic-Zunic}, {Baratoux}, {Baroukh}, {Barraclough}, {Battalio},
  {Beach}, {Bean}, {Beck}, {Becker}, {Beegle}, {Behar}, {Belgacem}, {Bell},
  {Bender}, {Benna}, {Bentz}, {Berger}, {Berger}, {Berlanga}, {Berman}, {Bish},
  {Blacksberg}, {Blake}, {Jos{\'e} Blanco}, {Blaney}, {Blank}, {Blau},
  {Bleacher}, {Boehm}, {Bonnet}, {Botta}, {B{\"o}ttcher}, {Boucher}, {Bower},
  {Boyd}, {Boynton}, {Braswell}, {Breves}, {Bridges}, {Bridges},
  {Brinckerhoff}, {Brinza}, {Bristow}, {Brunet}, {Brunner}, {Brunner}, {Buch},
  {Bullock}, {Burmeister}, {Burton}, {Buz}, {Cabane}, {Calef}, {Cameron},
  {Campbell}, {Cantor}, {Caplinger}, {Clifton}, {Caride Rodr{\'\i}guez},
  {Carmosino}, {Carrasco Bl{\'a}zquez}, {Cavanagh}, {Charpentier}, {Chipera},
  {Choi}, {Christensen}, {Clark}, {Clegg}, {Cleghorn}, {Cloutis}, {Cody},
  {Coll}, {Coman}, {Conrad}, {Coscia}, {Cousin}, {Cremers}, {Crisp}, {Cropper},
  {Cros}, {Cucinotta}, {d'Uston}, {Davis}, {Day}, {Daydou}, {DeFlores},
  {Dehouck}, {Delapp}, {DeMarines}, {Dequaire}, {Des Marais}, {Desrousseaux},
  {Dietrich}, {Dingler}, {Domagal-Goldman}, {Donny}, {Downs}, {Drake},
  {Dromart}, {Dupont}, {Duston}, {Dworkin}, {Dyar}, {Edgar}, {Edgett},
  {Edwards}, {Edwards}, {Edwards}, {Ehlmann}, {Ehresmann}, {Eigenbrode},
  {Elliott}, {Elliott}, {Ewing}, {Fabre}, {Fair{\'e}n}, {Fair{\'e}n}, {Farley},
  {Farmer}, {Fassett}, {Favot}, {Fay}, {Fedosov}, {Feldman}, {Fendrich},
  {Fischer}, {Fisk}, {Fitzgibbon}, {Flesch}, {Floyd}, {Fl{\"u}ckiger}, {Forni},
  {Fox}, {Fraeman}, {Francis}, {Fran{\c{c}}ois}, {Franz}, {Freissinet},
  {French}, {Frydenvang}, {Garvin}, {Gasnault}, {Geffroy}, {Gellert}, {Genzer},
  {Getty}, {Glavin}, {Godber}, {Goesmann}, {Goetz}, {Golovin}, {G{\'o}mez
  G{\'o}mez}, {G{\'o}mez-Elvira}, {Gondet}, {Gordon}, {Gorevan}, {Graham},
  {Grant}, {Grinspoon}, {Grotzinger}, {Guillemot}, {Guo}, {Gupta}, {Guzewich},
  {Haberle}, {Halleaux}, {Hallet}, {Hamilton}, {Hand}, {Hardgrove}, {Hardy},
  {Harker}, {Harpold}, {Harri}, {Harshman}, {Hassler}, {Haukka}, {Hayes},
  {Herkenhoff}, {Herrera}, {Hettrich}, {Heydari}, {Hipkin}, {Hoehler},
  {Hollingsworth}, {Hudgins}, {Huntress}, {Hurowitz}, {Hviid}, {Iagnemma},
  {Indyk}, {Isra{\"e}l}, {Jackson}, {Jacob}, {Jakosky}, {Jean-Rigaud},
  {Jensen}, {Kl{\o}vgaard Jensen}, {Johnson}, {Johnson}, {Johnstone}, {Jones},
  {Jones}, {Joseph}, {Joulin}, {Jun}, {Kah}, {Kahanp{\"a}{\"a}}, {Kahre},
  {Kaplan}, {Karpushkina}, {Kashyap}, {Kauhanen}, {Keely}, {Kelley}, {Kempe},
  {Kemppinen}, {Kennedy}, {Keymeulen}, {Kharytonov}, {Kim}, {Kinch}, {King},
  {Kirk}, {Kirkland}, {Kloos}, {Kocurek}, {Koefoed}, {K{\"o}hler}, {Kortmann},
  {Kotrc}, {Kozyrev}, {Krau}, {Krezoski}, {Kronyak}, {Krysak}, {Kuzmin},
  {Lacour}, {Lafaille}, {Langevin}, {Lanza}, {Lap{\^o}tre}, {Larif}, {Lasue},
  {Le Deit}, {Le Mou{\'e}lic}, {Lee}, {Lee}, {Lee}, {Lees}, {Lefavor},
  {Lemmon}, {Lepinette}, {Lepore}, {Leshin}, {L{\'e}veill{\'e}}, {Lewin},
  {Lewis}, {Li}, {Lichtenberg}, {Lipkaman}, {Lisov}, {Little}, {Litvak}, {Liu},
  {Lohf}, {Lorigny}, {Lugmair}, {Lundberg}, {Lyness}, {Madsen}, {Magee},
  {Mahaffy}, {Maki}, {M{\"a}kinen}, {Malakhov}, {Malespin}, {Malin}, {Mangold},
  {Manhes}, {Manning}, {Marchand}, {Mar{\'\i}n Jim{\'e}nez}, {Mart{\'\i}n
  Garc{\'\i}a}, {Martin}, {Martin}, {Martin}, {Mart{\'\i}nez Mart{\'\i}nez},
  {Mart{\'\i}nez-Fr{\'\i}as}, {Mart{\'\i}n-Sauceda}, {Mart{\'\i}n-Soler},
  {Mart{\'\i}n-Torres}, {Mason}, {Matthews}, {Matthi{\"a}}, {Mauchien},
  {Maurice}, {McAdam}, {McBride}, {McCartney}, {McConnochie}, {McCullough},
  {McEwan}, {McKay}, {McLain}, {McLennan}, {McNair}, {Melikechi}, {Mendaza de
  Cal}, {Merikallio}, {Merritt}, {Meslin}, {Meyer}, {Mezzacappa}, {Milkovich},
  {Millan}, {Miller}, {Miller}, {Milliken}, {Ming}, {Minitti}, {Mischna},
  {Mitchell}, {Mitrofanov}, {Moersch}, {Mokrousov}, {Molina}, {Moore},
  {Moores}, {Mora-Sotomayor}, {Moreno}, {Morookian}, {Morris}, {Morrison},
  {Mousset}, {Mrigakshi}, {Mueller-Mellin}, {Muller}, {Mu{\~n}oz Caro},
  {Nachon}, {Nastan}, {Navarro L{\'o}pez}, {Navarro Gonz{\'a}lez}, {Nealson},
  {Nefian}, {Nelson}, {Newcombe}, {Newman}, {Newsom}, {Nikiforov}, {Nikitczuk},
  {Niles}, {Nixon}, {Noblet}, {Noe}, {Nolan}, {Oehler}, {Ollila}, {Olson},
  {Orthen}, {Owen}, {Ozanne}, {de Pablo Hern{\'a}ndez}, {Pagel}, {Paillet},
  {Pallier}, {Palucis}, {Parker}, {Parot}, {Parra}, {Patel}, {Paton},
  {Paulsen}, {Pavlov}, {Pavri}, {Peinado-Gonz{\'a}lez}, {Pepin}, {Peret},
  {P{\'e}rez}, {Perrett}, {Peterson}, {Pilorget}, {Pinet}, {Pinnick},
  {Pla-Garc{\'\i}a}, {Plante}, {Poitrasson}, {Polkko}, {Popa}, {Posiolova},
  {Posner}, {Pradler}, {Prats}, {Prokhorov}, {Raaen}, {Radziemski}, {Rafkin},
  {Ramos}, {Rampe}, {Rapin}, {Raulin}, {Ravine}, {Reitz}, {Ren}, {Renn{\'o}},
  {Rice}, {Richardson}, {Ritter}, {Rivera-Hern{\'a}ndez}, {Robert},
  {Robertson}, {Rodriguez Manfredi}, {Jos{\'e} Romeral-Planell{\'o}},
  {Rowland}, {Rubin}, {Saccoccio}, {Said}, {Salamon}, {Sanin}, {Sans Fuentes},
  {Saper}, {Sarrazin}, {Sautter}, {Savij{\"a}rvi}, {Schieber}, {Schmidt},
  {Schmidt}, {Scholes}, {Schoppers}, {Schr{\"o}der}, {Schwenzer}, {Sciascia
  Borlina}, {Scodary}, {Sebasti{\'a}n Mart{\'\i}nez}, {Sengstacken}, {Shechet},
  {Shterts}, {Siebach}, {Siili}, {Simmonds}, {Sirven}, {Slavney}, {Sletten},
  {Smith}, {Sobron Sanchez}, {Spanovich}, {Spray}, {Spring}, {Squyres},
  {Stack}, {Stalport}, {Starr}, {Stein}, {Stern}, {Stewart}, {Stewart},
  {Stipp}, {Stoiber}, {Stolper}, {Sucharski}, {Sullivan}, {Summons}, {Sumner},
  {Sun}, {Supulver}, {Sutter}, {Szopa}, {Tan}, {Tate}, {Teinturier}, {ten
  Kate}, {Thomas}, {Thomas}, {Thompson}, {Thuillier}, {Thulliez}, {Tokar},
  {Toplis}, {de la Torre Ju{\'a}rez}, {Torres Redondo}, {Trainer}, {Treiman},
  {Tretyakov}, {Ull{\'a}n-Nieto}, {Urqui-O'Callaghan}, {Valent{\'\i}n-Serrano},
  {Van Beek}, {Van Beek}, {VanBommel}, {Vaniman}, {Varenikov}, {Vasavada},
  {Vasconcelos}, {de Vicente-Retortillo Rubalcaba}, {Vicenzi}, {Vostrukhin},
  {Voytek}, {Wadhwa}, {Ward}, {Watkins}, {Webster}, {Weigle}, {Wellington},
  {Westall}, {Wiens}, {Wilhelm}, {Williams}, {Williams}, {Williams},
  {Williams}, {Williford}, {Wilson}, {Wilson}, {Wimmer-Schweingruber}, {Wolff},
  {Wong}, {Wray}, {Yana}, {Yen}, {Yingst}, {Zeitlin}, {Zimdar}, \& {Zorzano
  Mier}}]{2015Sci...347..415W}
{Webster}, C.~R., {Mahaffy}, P.~R., {Atreya}, S.~K., {et~al.} 2015, Science,
  347, 415, \dodoi{10.1126/science.1261713}

\bibitem[{{Werthimer} {et~al.}(2001){Werthimer}, {Anderson}, {Bowyer}, {Cobb},
  {Heien}, {Korpela}, {Lampton}, {Lebofsky}, {Marcy}, {McGarry}, \&
  {Treffers}}]{2001SPIE.4273..104W}
{Werthimer}, D., {Anderson}, D., {Bowyer}, C.~S., {et~al.} 2001, in Society of
  Photo-Optical Instrumentation Engineers (SPIE) Conference Series, Vol. 4273,
  The Search for Extraterrestrial Intelligence (SETI) in the Optical Spectrum
  III, ed. S.~A. {Kingsley} \& R.~{Bhathal}, 104--109,
  \dodoi{10.1117/12.435384}

\bibitem[{Williams \& Pollard(2002)}]{williams_pollard_2002}
Williams, D.~M., \& Pollard, D. 2002, International Journal of Astrobiology, 1,
  61–69, \dodoi{10.1017/S1473550402001064}

\bibitem[{{Zhang} {et~al.}(2020){Zhang}, {Werthimer}, {Zhang}, {Cobb},
  {Korpela}, {Anderson}, {Gajjar}, {Lee}, {Li}, {Pei}, {Zhang}, {Huang},
  {Wang}, {Zhu}, {Duan}, {Zhang}, {Jin}, {Zhu}, \& {Li}}]{2020ApJ...891..174Z}
{Zhang}, Z.-S., {Werthimer}, D., {Zhang}, T.-J., {et~al.} 2020, \apj, 891, 174,
  \dodoi{10.3847/1538-4357/ab7376}

\bibitem[{Zhu {et~al.}(2020)Zhu, Li, Luo, Miao, Zhang, Spitler, Lorimer,
  Kramer, Champion, Yue, Cameron, Cruces, Duan, Feng, Han, Hobbs, Niu, Niu,
  Pan, Qian, Shi, Tang, Wang, Wang, Yuan, Zhang, Zhang, Cao, Feng, Gan, Gao,
  Gu, Guo, Hao, Huang, Huang, Jiang, Jin, Li, Li, Li, Liu, Pan, Peng, Qian,
  Shi, Song, Song, Sun, Sun, Wang, Wang, Wang, Xie, Yan, Yang, Yang, Yao, Yu,
  Yu, Zhang, Zhang, Zhang, Zheng, Zhou, Zhu, Zhu, Zhu, Zhu, \& Zhu}]{Zhu_2020}
Zhu, W., Li, D., Luo, R., {et~al.} 2020, The Astrophysical Journal, 895, L6,
  \dodoi{10.3847/2041-8213/ab8e46}

\end{thebibliography}
\bibliographystyle{aasjournal}

\end{document}